\documentclass[aps, prl, reprint, superscriptaddress, floatfix, amsmath,amssymb]{revtex4-2}
\pdfoutput=1

\usepackage[margin=5pt,caption=false]{subfig}
\usepackage{slantsc}
\usepackage{comment}
\usepackage{threeparttable}
\usepackage{booktabs}
\usepackage[figuresright]{rotating}
\usepackage{textcase}
\usepackage{xspace} 
\usepackage{multirow}
\usepackage[dvipsnames]{xcolor} 
\usepackage{booktabs} 
\usepackage{float}

\newcommand{\AS}{{AbacusSummit}\xspace}
\newcommand{\FM}{{\sc flamingo}\xspace}

\newcommand{\hiMsun}{h^{-1}M_{\odot}}
\newcommand{\hiMpc}{h^{-1}\mathrm{Mpc}}
\newcommand{\wpgg}{w_{p,\mathrm{gg}}}
\newcommand{\wpcg}{w_{p,\mathrm{cg}}}
\newcommand{\rp}{r_{p}}
\newcommand{\DSg}{\Delta\Sigma_{g}}
\newcommand{\DSc}{\Delta\Sigma_{c}}
\newcommand{\ngal}{n_{g}}

\newcommand{\om}{$\Omega_{m}$\xspace}
\newcommand{\Om}{\Omega_{m}}

\newcommand{\Ob}{\Omega_{b}}

\newcommand{\CxG}{C$\times$G\xspace}

\usepackage{array}
\usepackage[T1]{fontenc}
\usepackage{scalerel}
\usepackage{tikz}
\usetikzlibrary{svg.path}

\definecolor{orcidlogocol}{HTML}{A6CE39}
\tikzset{
  orcidlogo/.pic={
    \fill[orcidlogocol] svg{M256,128c0,70.7-57.3,128-128,128C57.3,256,0,198.7,0,128C0,57.3,57.3,0,128,0C198.7,0,256,57.3,256,128z};
    \fill[white] svg{M86.3,186.2H70.9V79.1h15.4v48.4V186.2z}
                 svg{M108.9,79.1h41.6c39.6,0,57,28.3,57,53.6c0,27.5-21.5,53.6-56.8,53.6h-41.8V79.1z M124.3,172.4h24.5c34.9,0,42.9-26.5,42.9-39.7c0-21.5-13.7-39.7-43.7-39.7h-23.7V172.4z}
                 svg{M88.7,56.8c0,5.5-4.5,10.1-10.1,10.1c-5.6,0-10.1-4.6-10.1-10.1c0-5.6,4.5-10.1,10.1-10.1C84.2,46.7,88.7,51.3,88.7,56.8z};
  }
}

\newcommand\orcidicon[1]{\href{https://orcid.org/#1}{\mbox{\scalerel*{
\begin{tikzpicture}[yscale=-1,transform shape]
\pic{orcidlogo};
\end{tikzpicture}
}{|}}}}

\usepackage{amsmath,bm}	
\usepackage{amssymb}	
\usepackage{fnpct}

\usepackage[colorlinks=true,allcolors=blue]{hyperref}
\begin{document}


\title{Unifying cluster and galaxy cosmology analyses using the galaxy-halo connection 
}

\author{Shulei Cao$^{\orcidicon{0000-0003-2421-7071}}$}
 \email[Email: ]{shuleic@mail.smu.edu}
\affiliation{Department of Physics, Southern Methodist University, Dallas, TX 75205, USA}

\author{Hao-Yi Wu$^{\orcidicon{0000-0002-7904-1707}}$}
\email[Email: ]{hywu@mail.smu.edu}
\affiliation{Department of Physics, Southern Methodist University, Dallas, TX 75205, USA}

\author{Andr\'es~N.~Salcedo$^{\orcidicon{0000-0003-1420-527X}}$}
\affiliation{Department of Astronomy/Steward Observatory, University of Arizona, Tucson, AZ 85721, USA}
\affiliation{Department of Physics, University of Arizona, Tucson, AZ 85721, USA}

\author{David~H.~Weinberg$^{\orcidicon{0000-0001-7775-7261}}$}
\affiliation{Department of Astronomy and Center for Cosmology and AstroParticle Physics (CCAPP), The Ohio State University, Columbus, OH 43210, USA}

\author{Matthieu Schaller$^{\orcidicon{0000-0002-2395-4902}}$}
\affiliation{Leiden Observatory, Leiden University, PO Box 9513, 2300 RA Leiden, the Netherlands}
\affiliation{Lorentz Institute for Theoretical Physics, Leiden University, PO Box 9506, 2300 RA Leiden, the Netherlands}

\author{Joop Schaye$^{\orcidicon{0000-0002-0668-5560}}$}
\affiliation{Leiden Observatory, Leiden University, PO Box 9513, 2300 RA Leiden, the Netherlands}


\date{\today}

\begin{abstract}
Galaxies and galaxy clusters trace the same cosmic density field, but their statistics have been modeled separately in cosmological analyses.  We present a unified, simulation-based framework to model them using the galaxy-halo connection.  Our analysis includes cluster lensing, galaxy clustering, and galaxy-cluster cross-correlation. We validate our method on the {\sc flamingo} hydrodynamic simulation. 
Relative to the cluster-only approach, combining these probes improves the $\sigma_8-\Omega_m$ figure of merit by a factor of 15. Our framework enables stringent tests of cosmological models and exploits small-scale information.
\end{abstract}
\maketitle

\emph{Introduction}---The standard model of cosmology built on a cosmological constant and cold dark matter ($\Lambda$CDM), after passing decades of rigorous experimental tests, is now being challenged by various tensions \cite{Verde19, Abdalla22, Perivolaropoulos22}.    
Besides the Hubble tension, an evolving dark energy equation of state has been evoked by the recent result from the Dark Energy Spectroscopic Instrument (DESI) \cite{DESI24BAOcosmo, DESI25BAOcosmo}, which corresponds to the $\Om$ tension (the matter density parameter inferred from baryon acoustic oscillations is lower than that from supernovae \cite{Brout22Pantheon+, DESI24BAOcosmo, DES25BAO+SN, Salcedo2025b}).
In addition, early-universe probes tend to infer a higher density fluctuation parameter $\sigma_8$ than late-universe probes \cite{Planck18cosmo, Li22HSC, Dalal23HSC, AmonRobertson23}.
All these tensions could indicate new physics or unaddressed systematics in our experiments.

Wide-field optical imaging surveys---including the Dark Energy Survey (DES), the Vera Rubin Observatory Legacy Survey of Space and Time (LSST), and the Roman Space Telescope---simultaneously detect galaxies, galaxy clusters, and gravitational lensing shear, holding the promise of pinning down $\Om$, $\sigma_8$, and dark energy by measuring the growth of structure. 
To maximize the constraining power on cosmological parameters, these probes should be combined optimally and self-consistently.  However, two major hurdles have made it difficult to combine the small-scale information of these probes \citep{Eifler21, DESY14x2ptN, DESY3CL+3x2pt, To25}.
The first hurdle is that clusters and galaxies have been modeled very differently. 
Cluster probes---cluster counts and lensing---are usually modeled using an observable-mass relation \cite{LimaHu2005, DESY1CL, Salcedoetal2020, To25}, while galaxy probes---galaxy clustering and galaxy-galaxy lensing---are usually modeled using galaxy bias or the galaxy-halo connection \citep{WechslerTinker18, Zacharegkas21, DESY3Magnification2x2pt, KiDS2x2pt+SMF, HSCY32x2pt}. Thus, the combined analyses tend to have a large number of astrophysical parameters, which weakens the constraining power \citep{Eifler21, DESY14x2ptN, To25, DESY3CL+3x2pt}.  The second hurdle is that line-of-sight structure biases optical cluster selection, which in turn affects mass calibration and cosmological results \citep{DESY1CL, Sunayama20, Wuetal2022, Zhouetal2024, Salcedoetal2024, Cao25association, Ndeetal2025}. Current mitigation strategies include removing small scales and adding extra model parameters, which further weakens the constraining power \citep{Zeng23, Parketal2023, Sunayamaetal2024, To25}. 

To overcome these hurdles, we propose a unified framework for modeling clusters and galaxies, building on our previous work \citep{Wibkingetal2020, Salcedoetal2020, Salcedoetal2022, Zeng23, Salcedoetal2024, Leeetal2024}.  We use the galaxy-halo connection to describe both galaxies and clusters, modeling cluster projection effects together with galaxy clustering.  We no longer explicitly perform cluster mass calibration; the cluster observable-mass relation becomes a prediction rather than an empirical model.  This approach significantly reduces the parameter space and unlocks the constraining power from small scales.  Further, we can self-consistently model the small-scale cluster-galaxy cross-correlation function.

\emph{A unified framework for clusters and galaxies}---The key to our unified framework is using the same galaxy sample for both galaxy analyses and cluster finding. We first select a luminous red galaxy sample for the galaxy clustering analysis, and we perform cluster finding using the same galaxy sample.  
To model this galaxy sample, we apply a halo occupation distribution (HOD) model to $N$-body simulations to generate a mock galaxy catalog, and we perform cluster finding on these galaxies to generate a cluster catalog.  We focus on the following statistics: 
\begin{itemize}
    \item Cluster lensing (CL): $\DSc$, which implicitly includes cluster number counts through number density matching \cite{Umetsu20, Salcedoetal2024}
    \item Galaxy number density and clustering (GG): $\ngal$ and $\wpgg$ 
    \item Cluster-galaxy cross-correlation (\CxG): $\wpcg$ 
    \item Galaxy-galaxy lensing (GGL): $\DSg$ 
\end{itemize}

The weak lensing signal is related to the excess surface mass density via
\begin{equation}
    \Delta\Sigma(\rp) = \overline{\Sigma}(<\rp) - \Sigma(\rp),
\end{equation}
where $\Sigma(\rp)$ is the surface density at projected radius $\rp$, and $\overline{\Sigma}(<\rp)$ is the mean surface density inside $\rp$. 
The projected correlation function is given by
\begin{equation}
    w_{p}(\rp)=2\int^{\pi_{\rm max}}_0\xi(\rp,\pi)d\pi,
\end{equation}
where $\xi$ is the anisotropic two-point correlation function depending on the projected separation $\rp$ and the line-of-sight separation $\pi$.  We use \texttt{Corrfunc} \citep{CORRFUNC1} to compute these data vectors.
We calculate the covariance matrices following \cite{Wuetal2019, Salcedoetal2020} using the survey assumptions listed in Table~\ref{tab:survey_parameters}.

\emph{Implementation with \AS}---We implement the unified modeling framework using the \AS $N$-body simulation suite \cite{ABACUSSUMMIT}.
Using the $z=0.3$ snapshot for demonstration, we build mock galaxy and cluster catalogs across a wide range of cosmological and astrophysical parameters, calculate the data vectors, and interpolate the data vectors as a function of these parameters.

The emulator suite of \AS includes 52 different cosmologies. Each simulation has a box size of $2~h^{-1}\text{Gpc}$, a mass resolution of $\sim2\times 10^9~h^{-1}M_{\odot}$, and a force resolution of 7.2 $h^{-1}$ proper kpc. 
We use the halos identified by the {CompaSO} halo finder and cleaned with merger trees \citep{HadzhiyskaCOMPASO, BoseMergerTrees}, using the L1 halo masses defined with virial spherical overdensity \cite{BryanNorman1998}.

We first assign galaxies to dark matter halos using a simple HOD model motivated by \citep{Zheng07, Salcedoetal2022}.  The mean numbers of central and satellite galaxies are given by
\begin{align}
\nonumber
    \phantom{=}\langle N_{\text{cen}} | M_{h} \rangle &= \frac{f_{\text{cen}}}{2} \left[ 1 + \text{erf} \left( \frac{\log_{10} M_{h} - \log_{10} M_{\text{min}}}{\sigma_{\log_{10} M}} \right) \right], \\
    \phantom{=}\langle N_{\text{sat}} | M_{h} \rangle &= \frac{\langle N_{\text{cen}} | M_{h} \rangle}{f_{\text{cen}}} \left( \frac{M_{h} - M_{0}}{M_{1}} \right)^{\alpha},\label{HOD_model}
\end{align}
where $M_{h}$ is halo mass, $M_{\text{min}}$ is the mass threshold for hosting a central galaxy, $\sigma_{\log_{10} M}$ is the width of transitioning from zero to one central galaxy, $M_{0}$ is the cutoff mass for hosting a satellite, $M_{1}$ is the characteristic mass for hosting a satellite, and $\alpha$ determines how satellite number scales with halo mass.
The spatial distribution of satellites follows an NFW profile \cite{NFW1997}. The galaxy assembly bias is based on \cite{Salcedoetal2022}.

We then perform cluster finding on these galaxies, counting the number of galaxies ($N_{\rm cyl}$) within a cylinder of radius $R_{\rm cyl} = (N_{\rm cyl}/100)^{0.2}\,\hiMpc$ and line-of-sight distance $\pm 90\,\hiMpc$ around each halo center \citep{Salcedoetal2024, Leeetal2024}. 
Here $N_{\rm cyl}$ and $R_{\rm cyl}$ are determined iteratively.
To avoid double-counting galaxies in overlapping cylinders, we preferentially associate galaxies with more massive clusters \citep{Rykoffetal2014}. 

In observations, $w_p$ and $\Delta\Sigma$ are computed assuming a fiducial cosmology. To mimic this process, when we measure $w_p$ and $\Delta\Sigma$ using mock catalogs, we use a fiducial cosmology ($\Om=0.3$ and $h=0.7$) to convert angular and redshift separations into transverse and line-of-sight distances.  We also use the fiducial cosmology to compute the lensing critical density \cite{Salcedo2025a}.

For each of the 52 cosmologies, we select 100 sets of HOD parameters via a hypersphere sampling \citep{Nygaardetal2024}, limited to models yielding galaxy densities within a factor of two of the observed galaxy density. With these 5200 simulations, we build an emulator to interpolate the data vector as a function of model parameters, following the Gaussian process regression method described in \cite{Wibkingetal2020}.  The emulator uncertainties are significantly smaller than the error bars expected from our survey assumptions.

\begin{table}
\centering
\setlength\tabcolsep{6.5pt}
\begin{threeparttable}
\caption{Survey parameters assumed in the forecast.  We assume the DES Y1 source galaxy distribution \cite{DESY1CL}. 
}
\begin{tabular}{l|l}
\toprule\toprule
Survey volume  & 1.0 Gpc$^3$ \\
Equivalent survey coverage & 4400 deg$^2, z\in[0.2,0.35)$ \\
\midrule
Cluster number density &  $1.4\times10^{-5}~\hiMpc^{-3}$ \\
Lens galaxy number density &   $6.0\times10^{-3}~\hiMpc^{-3}$ \\
\midrule
Total number of clusters  &  4300 \\
Total number of lens galaxies  &  $1.9\times10^6$ \\
\bottomrule\bottomrule
\end{tabular}
\label{tab:survey_parameters}
\end{threeparttable}%
\end{table}

\begin{figure*}
\centering
\includegraphics[width=2\columnwidth]{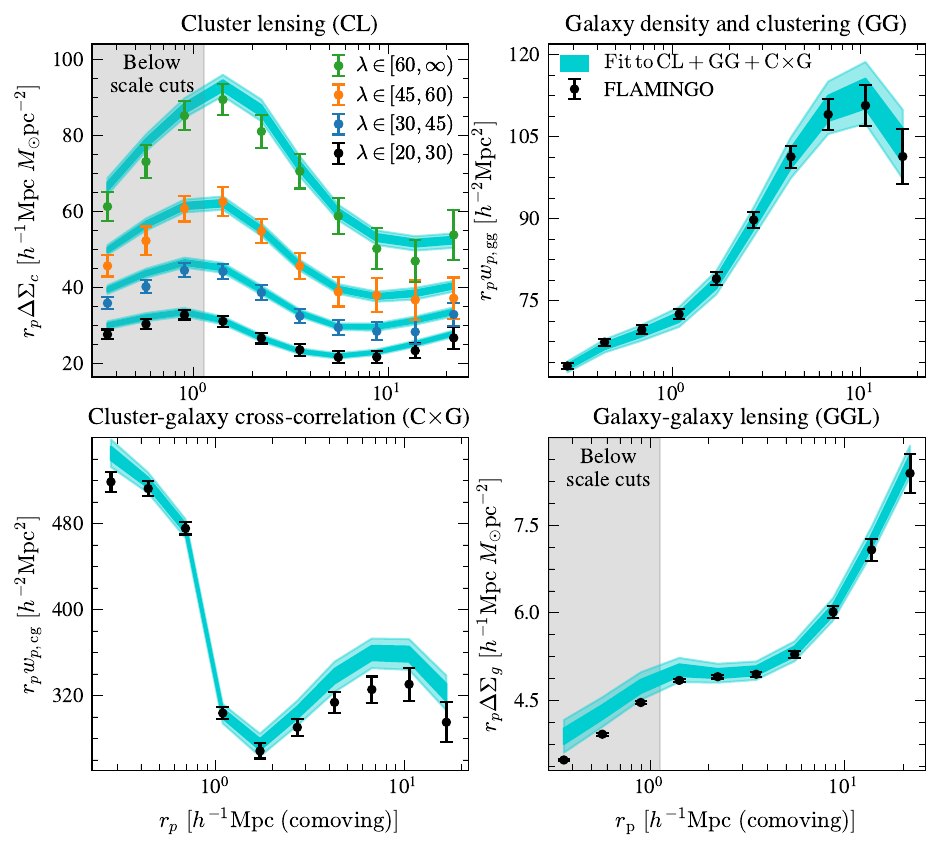}
\caption{Validation of our unified framework.
The points with error bars correspond to the mock observations from \FM, and the four panels correspond to cluster lensing (CL), galaxy density and clustering (GG), cluster-galaxy cross-correlation (\CxG), and galaxy-galaxy lensing (GGL). We fit our models to the first three (excluding GGL due to uncertainties in baryonic physics), and the bands correspond to the posterior predictions.  Our model describes all these data vectors well and leads to unbiased, competitive cosmological constraints (see Fig.~\ref{fig:mcmc}).
}
\label{fig:data_vectors}
\end{figure*}

\begin{table*}[htbp!]
\centering
\setlength\tabcolsep{6pt}
\begin{threeparttable}
\caption{Forecasted constraining power for various data vector combinations.  The figure of merit is defined as 
${\rm FoM}=10^{-3}[\det{\rm Cov}(\sigma_8, \Om)]^{-1/2}$.
The non-unified approach improves upon the cluster-only case (CL) by a factor of 2.6, while the unified approach improves the FoM by a factor of 9.
Including cluster-galaxy cross-correlation (\CxG) further improves the FoM by a factor of 1.7. Adding galaxy-galaxy lensing (GGL) improves the FoM by an additional factor of 1.8. 
}
\label{tab:FoM}
\begin{tabular}{ll|llll|llll}
\toprule\toprule
 & & \multicolumn{4}{c|}{Constraints without GGL} & \multicolumn{4}{c}{Constraints with GGL ($\DSg$)} \\
 & Data vector & $\sigma_8$ & \om & $S_8$ & FoM & $\sigma_8$ & \om & $S_8$ & FoM \\
\midrule
CL & $\DSc$ & 6.5\%  & 12.8\% & 3.9\% & 0.8 & & & & \\[2pt]
GG & $\wpgg + \ngal$ & 4.8\% & 5.6\% & 5.8\% & 1.2 & 2.4\% & 4.5\% & 0.8\% & 11.0 \\[2pt]
CL + GG (non-unified) & $\DSc + (\wpgg + \ngal)$ & 3.5\% & 5.5\% & 3.8\% & 2.1 & 2.2\% & 4.3\% & 0.8\% & 12.2 \\[2pt]
CL + GG (unified) & same as above & 2.5\% & 5.7\% & 1.1\% & {\bf 7.4} & 2.1\% & 4.3\% & 0.6\% & 15.0 \\[2pt]
CL + GG + \CxG (unified) & $\DSc + (\wpgg + \ngal) + \wpcg$ & 1.6\% & 3.7\% & 1.0\% & {\bf 12.3} & 1.5\% & 3.1\% & 0.6\% & {\bf 22.7} \\
\bottomrule\bottomrule
\end{tabular}
\end{threeparttable}%
\end{table*}

\emph{Mock observations with \FM}---We would like to demonstrate that our unified framework has the flexibility to describe realistic galaxy samples.
To this end, we fit our model to the data vector generated from the \FM hydrodynamic simulations, with a baryonic feedback model tuned to match the observed stellar mass function and cluster gas mass fraction \cite{Schayeetal2023, Kugel23Flamingo, Kugel25FlamingoCluster}.  The 1~Gpc$^3$ volume of \FM is much larger than previous hydrodynamic simulations of similar resolution, allowing us to validate our model at precision comparable to current surveys.

We use the highest-resolution box with galaxies identified by the {\tt HBT-Herons} algorithm \citep{Hanetal2012,Moreno2025HBT}.  To select luminous red galaxies, we first fit the red-sequence color-magnitude relation using galaxies in halos with virial mass from the SOAP (Spherical Overdensity and Aperture Processor) catalog, $M_{h}> 5\times10^{14}~\hiMsun$.  Galaxies are selected with a color-space $\chi^2 < 6$ and a $z$-band magnitude cut to have a density of $6 \times 10^{-3}~h^3\text{Mpc}^{-3}$, motivated by the DES redMaGiC sample with a broad color cut \cite{Rozoetal2016}.   With this galaxy sample, we construct a cluster sample using counts-in-cylinders, with 4 richness bins defined by matching the cluster number densities in the DES Year 1 sample \cite{DESY1CL}. We remove $\rp$ bins below $1.3$ physical Mpc to avoid the impact of baryonic feedback at small scales.

\emph{Mock likelihood analysis}---We fit our model to the \FM data vector, with covariance matrices calculated with the survey condition listed in Table~\ref{tab:survey_parameters}. 
Our Markov chain Monte Carlo \citep{emcee} has 8 HOD parameters ($\log M_{\rm min}, \sigma_{\log M}, f_{\rm cen}, \alpha, M_0/M_{1}, M_1/M_{\rm min}, Q_{\rm cen}, Q_{\rm sat}$) and 4 cosmological parameters ($\Om$, $\sigma_8$, $\Ob$, $n_{s}$), with flat priors set by the boundaries of the emulator grid.
Our emulator inherits a fixed CMB acoustic scale $\theta_*$ imposed in \AS, which sets $h$. Since \FM does not share this prior, the $\Ob$ and $n_s$ constraints are mildly shifted, but the constraints $\Om$, $\sigma_8$ are unaffected.

Figure~\ref{fig:data_vectors} shows that our posterior prediction (bands) well describes the full \FM data vectors (points with error bars).  
We exclude GGL in  the fitting because it is more affected by baryonic effects compared to cluster lensing \cite{Leauthaud17, Lange19, Krause21, AmonRobertson23}, but the posterior predictions of our model agree with GGL, as shown in the bottom-right panel.

\begin{figure}
\centering
\includegraphics[width=\columnwidth]{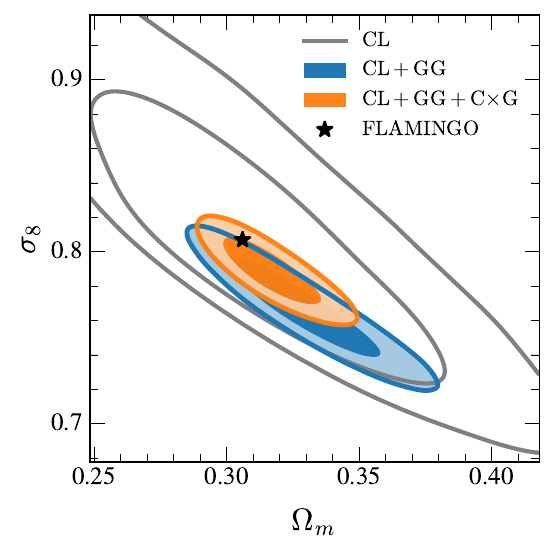}
\caption{Forecasted constraints derived from fitting the unified model to simulated data vector combinations.  With the unified framework, adding galaxy clustering (GG) significantly improves the constraints from cluster lensing alone (CL).  Adding cluster-galaxy cross correlation (\CxG) further tightens the constraints (also see Table~\ref{tab:FoM}).  We assume a survey volume matching the \FM simulation, 1 Gpc$^3$.}
\label{fig:mcmc}
\end{figure}

\emph{Forecasted cosmological constraints}---Figure~\ref{fig:mcmc} shows the $\sigma_8-\Om$ constraints from three of our mock analyses, where the simulation's true values are recovered at $0.48\sigma$, $2.8\sigma$, and $1.6\sigma$.
From cluster-only (CL) to adding galaxy clustering (CL+GG), the contours shrink significantly.  Adding the cluster-galaxy cross-correlation (\CxG) further shrinks the contours.

Table~\ref{tab:FoM} presents the parameter constraints from various combinations of data vectors.  We present the 68\% intervals for $\sigma_8$, $\Om$, and $S_8=\sigma_8\sqrt{\Om/0.3}$, along with the figure of merit of $\sigma_8-\Om$, defined as ${\rm FoM}=10^{-3}[\det{\rm Cov}(\sigma_8, \Om)]^{-1/2}$.  
We present results including or excluding GGL because it is more sensitive to baryonic effects than other probes \cite{Leauthaud17, Lange19, Krause21, AmonRobertson23}.
The first two rows show the CL-only and GG-only results. 
The third row combines CL and GG {\em without} unifying their modeling, i.e., we allow separate values of HOD parameters when fitting the two probes.
The improvement is modest because of the large number of free parameters.
This non-unified approach represents the state of the art in modeling clusters and galaxies separately \cite{Salcedoetal2020, DESY3CL+3x2pt, To25}.

Let us first focus on the constraints without GGL (the middle group of columns).
The fourth row shows that our unified approach can improve the FoM by a factor of 3.5 compared with non-unified (2.1 to 7.4), and by a factor of 9 compared with CL-only (0.8 to 7.4). This improvement is due to self-calibration of the shared galaxy-halo connection parameters.
The last row includes \CxG, which improves the FoM further by a factor of 1.7 (7.4 to 12.3; also see the orange contours in Fig.~\ref{fig:mcmc}).
We emphasize that a non-unified approach cannot model small-scale \CxG self-consistently, and previous \CxG analyses are limited to large scales \cite{To25, DESY3CL+3x2pt}.
The last group of columns adds the GGL information.  Without \CxG, the gain due to GGL is modest (FoM = 15.0); with \CxG, the gain due to GGL is significantly enhanced (FoM = 22.7).  Therefore, including both \CxG and GGL is the most promising route, provided we can keep GGL's systematics under control.

\emph{Summary and discussion}---We have presented a unified framework for modeling clusters (C) and galaxies (G) using $N$-body simulations populated with an HOD model.
Our framework can improve the cluster-only $\sigma_8-\Om$ FoM by a factor of 9 when adding GG, by a factor of 15 when further adding \CxG, and by a factor of 27.5 when further adding GGL.
Small-scale \CxG drives much of this gain, and our unified framework is the only viable approach for including small-scale \CxG in multi-probe studies.

Our forecast is based on a survey of 1 Gpc$^3$. The DES cluster sample has a volume of $\sim$ 7 Gpc$^3$ (assuming 5,000 deg$^2$ out to $z=0.65$), and the LSST cluster sample will have a volume of $\sim$ 60 Gpc$^3$ (assuming 15,000 deg$^2$ out to $z=1$).  The relative improvement between different probes should scale with the survey volume.

The \FM hydrodynamic simulations, with their complex astrophysics, validate our unified HOD framework. We plan to apply it to recent DES and upcoming LSST data. Validating the method at the precision achievable with these data sets will require simulations with volumes comparable to the surveys, which, for now, is only possible with $N$-body simulations. We will also need to account for observational systematics, including photometric redshift uncertainties \cite{Myles21SOMPZ} and shear bias \cite{Gatti21shear, Yamamoto25}. Nevertheless, our unified framework has overcome the key modeling hurdles in combining clusters and galaxies, and we expect it can fully realize the constraining power of wide-field imaging surveys.

\emph{Acknowledgments}---This work is supported by the DOE award DE-SC0010129 and the NSF award AST-2509910. ANS received funding from DOE grants DE-SC0009913, DE-SC0020247, and DE-SC0025993. We thank the \AS team for making their simulations publicly available.


\bibliographystyle{apsrev4-2-author-truncate}
\bibliography{apssamp}

\newcommand{\apjl}{Astrophys. J. Lett.}\newcommand{\apjs}{Astrophys. J. Suppl.}\newcommand{\mnras}{Mon. Not. R. Astron. Soc.}\newcommand{\jcap}{J. Cosmol. Astropart. Phys.}\newcommand{\aap}{Astron. Astrophys.}\newcommand{\rmxaa}{Revista Mexicana de Astronom{\'i}a y Astrof{\'i}sica}\newcommand{\phiv}{$\phi$}\newcommand{\apss}{Astrophys. Space Sci.}\newcommand{\pasp}{PASP}\newcommand{\nar}{New Astrono. Rev.}\newcommand{\pasj}{PASJ}\newcommand{\physrep}{Phys. Rep.}\newcommand{\aj}{Astron. J.}\newcommand{\araa}{Annu. Rev. Astron. Astrophys.}\newcommand{\aapr}{Astron. Astrophys. Rev.}\newcommand{\na}{Nat. Astron.}
\begin{thebibliography}{61}%
\makeatletter
\providecommand \@ifxundefined [1]{%
 \@ifx{#1\undefined}
}%
\providecommand \@ifnum [1]{%
 \ifnum #1\expandafter \@firstoftwo
 \else \expandafter \@secondoftwo
 \fi
}%
\providecommand \@ifx [1]{%
 \ifx #1\expandafter \@firstoftwo
 \else \expandafter \@secondoftwo
 \fi
}%
\providecommand \natexlab [1]{#1}%
\providecommand \enquote  [1]{``#1''}%
\providecommand \bibnamefont  [1]{#1}%
\providecommand \bibfnamefont [1]{#1}%
\providecommand \citenamefont [1]{#1}%
\providecommand \href@noop [0]{\@secondoftwo}%
\providecommand \href [0]{\begingroup \@sanitize@url \@href}%
\providecommand \@href[1]{\@@startlink{#1}\@@href}%
\providecommand \@@href[1]{\endgroup#1\@@endlink}%
\providecommand \@sanitize@url [0]{\catcode `\\12\catcode `\$12\catcode `\&12\catcode `\#12\catcode `\^12\catcode `\_12\catcode `\%12\relax}%
\providecommand \@@startlink[1]{}%
\providecommand \@@endlink[0]{}%
\providecommand \url  [0]{\begingroup\@sanitize@url \@url }%
\providecommand \@url [1]{\endgroup\@href {#1}{\urlprefix }}%
\providecommand \urlprefix  [0]{URL }%
\providecommand \Eprint [0]{\href }%
\providecommand \doibase [0]{https://doi.org/}%
\providecommand \selectlanguage [0]{\@gobble}%
\providecommand \bibinfo  [0]{\@secondoftwo}%
\providecommand \bibfield  [0]{\@secondoftwo}%
\providecommand \translation [1]{[#1]}%
\providecommand \BibitemOpen [0]{}%
\providecommand \bibitemStop [0]{}%
\providecommand \bibitemNoStop [0]{.\EOS\space}%
\providecommand \EOS [0]{\spacefactor3000\relax}%
\providecommand \BibitemShut  [1]{\csname bibitem#1\endcsname}%
\let\auto@bib@innerbib\@empty
\bibitem [{\citenamefont {{Verde}}\ \emph {et~al.}(2019)\citenamefont {{Verde}}, \citenamefont {{Treu}},\ and\ \citenamefont {{Riess}}}]{Verde19}%
  \BibitemOpen
  \bibfield  {author} {\bibinfo {author} {\bibfnamefont {L.}~\bibnamefont {{Verde}}}, \bibinfo {author} {\bibfnamefont {T.}~\bibnamefont {{Treu}}},\ \bibnamefont {and}\ \bibinfo {author} {\bibfnamefont {A.~G.}\ \bibnamefont {{Riess}}},\ }\href {https://doi.org/10.1038/s41550-019-0902-0} {\bibfield  {journal} {\bibinfo  {journal} {Nature Astronomy}\ }\textbf {\bibinfo {volume} {3}},\ \bibinfo {pages} {891} (\bibinfo {year} {2019})}\BibitemShut {NoStop}%
\bibitem [{\citenamefont {{Abdalla}}\ \emph {et~al.}(2022)\citenamefont {{Abdalla}}, \citenamefont {{Abell{\'a}n}}, \citenamefont {{Aboubrahim}}, \citenamefont {{Agnello}}, \citenamefont {{Akarsu}}, \citenamefont {{Akrami}}, \citenamefont {{Alestas}}, \citenamefont {{Aloni}}, \citenamefont {{Amendola}}, \citenamefont {{Anchordoqui}}, \citenamefont {{Anderson}}, \citenamefont {{Arendse}}, \citenamefont {{Asgari}} \emph {et~al.}}]{Abdalla22}%
  \BibitemOpen
  \bibfield  {author} {\bibinfo {author} {\bibfnamefont {E.}~\bibnamefont {{Abdalla}}}, \bibnamefont {et~al.},\ }\href {https://doi.org/10.1016/j.jheap.2022.04.002} {\bibfield  {journal} {\bibinfo  {journal} {J. High Energy Astrophys.}\ }\textbf {\bibinfo {volume} {34}},\ \bibinfo {pages} {49} (\bibinfo {year} {2022})}\BibitemShut {NoStop}%
\bibitem [{\citenamefont {{Perivolaropoulos}}\ and\ \citenamefont {{Skara}}(2022)}]{Perivolaropoulos22}%
  \BibitemOpen
  \bibfield  {author} {\bibinfo {author} {\bibfnamefont {L.}~\bibnamefont {{Perivolaropoulos}}}\ \bibnamefont {and}\ \bibinfo {author} {\bibfnamefont {F.}~\bibnamefont {{Skara}}},\ }\href {https://doi.org/10.1016/j.newar.2022.101659} {\bibfield  {journal} {\bibinfo  {journal} {\nar}\ }\textbf {\bibinfo {volume} {95}},\ \bibinfo {eid} {101659} (\bibinfo {year} {2022})}\BibitemShut {NoStop}%
\bibitem [{\citenamefont {{Adame}}\ \emph {et~al.}(2025)\citenamefont {{Adame}}, \citenamefont {{Aguilar}}, \citenamefont {{Ahlen}}, \citenamefont {{Alam}}, \citenamefont {{Alexander}}, \citenamefont {{Alvarez}}, \citenamefont {{Alves}}, \citenamefont {{Anand}}, \citenamefont {{Andrade}}, \citenamefont {{Armengaud}}, \citenamefont {{Avila}}, \citenamefont {{Aviles}}, \citenamefont {{Awan}}, \citenamefont {{Bahr-Kalus}}, \citenamefont {{Bailey}}, \citenamefont {{Baltay}}, \citenamefont {{Bault}}, \citenamefont {{Behera}}, \citenamefont {{BenZvi}}, \citenamefont {{Bera}}, \citenamefont {{Beutler}}, \citenamefont {{Bianchi}}, \citenamefont {{Blake}}, \citenamefont {{Blum}}, \citenamefont {{Brieden}}, \citenamefont {{Brodzeller}}, \citenamefont {{Brooks}}, \citenamefont {{Buckley-Geer}}, \citenamefont {{Burtin}}, \citenamefont {{Calderon}}, \citenamefont {{Canning}}, \citenamefont {{Carnero Rosell}}, \citenamefont {{Cereskaite}}, \citenamefont {{Cervantes-Cota}}, \citenamefont {{Chabanier}}, \citenamefont
  {{Chaussidon}}, \citenamefont {{Chaves-Montero}}, \citenamefont {{Chen}}, \citenamefont {{Chen}}, \citenamefont {{Claybaugh}}, \citenamefont {{Cole}}, \citenamefont {{Cuceu}}, \citenamefont {{Davis}}, \citenamefont {{Dawson}}, \citenamefont {{de la Macorra}}, \citenamefont {{de Mattia}}, \citenamefont {{Deiosso}}, \citenamefont {{Dey}}, \citenamefont {{Dey}}, \citenamefont {{Ding}}, \citenamefont {{Doel}}, \citenamefont {{Edelstein}}, \citenamefont {{Eftekharzadeh}}, \citenamefont {{Eisenstein}}, \citenamefont {{Elliott}}, \citenamefont {{Fagrelius}}, \citenamefont {{Fanning}}, \citenamefont {{Ferraro}}, \citenamefont {{Ereza}}, \citenamefont {{Findlay}}, \citenamefont {{Flaugher}}, \citenamefont {{Font-Ribera}}, \citenamefont {{Forero-S{\'a}nchez}}, \citenamefont {{Forero-Romero}}, \citenamefont {{Frenk}}, \citenamefont {{Garcia-Quintero}}, \citenamefont {{Gazta{\~n}aga}}, \citenamefont {{Gil-Mar{\'\i}n}}, \citenamefont {{Gontcho a Gontcho}}, \citenamefont {{Gonzalez-Morales}}, \citenamefont
  {{Gonzalez-Perez}}, \citenamefont {{Gordon}}, \citenamefont {{Green}}, \citenamefont {{Gruen}}, \citenamefont {{Gsponer}}, \citenamefont {{Gutierrez}}, \citenamefont {{Guy}}, \citenamefont {{Hadzhiyska}}, \citenamefont {{Hahn}}, \citenamefont {{Hanif}}, \citenamefont {{Herrera-Alcantar}}, \citenamefont {{Honscheid}}, \citenamefont {{Howlett}}, \citenamefont {{Huterer}}, \citenamefont {{Ir{\v{s}}i{\v{c}}}}, \citenamefont {{Ishak}}, \citenamefont {{Juneau}}, \citenamefont {{Kara{\c{c}}ayl{\i}}}, \citenamefont {{Kehoe}}, \citenamefont {{Kent}}, \citenamefont {{Kirkby}}, \citenamefont {{Kremin}}, \citenamefont {{Krolewski}}, \citenamefont {{Lai}}, \citenamefont {{Lan}}, \citenamefont {{Landriau}}, \citenamefont {{Lang}}, \citenamefont {{Lasker}}, \citenamefont {{Le Goff}}, \citenamefont {{Le Guillou}}, \citenamefont {{Leauthaud}}, \citenamefont {{Levi}}, \citenamefont {{Li}}, \citenamefont {{Linder}}, \citenamefont {{Lodha}}, \citenamefont {{Magneville}}, \citenamefont {{Manera}}, \citenamefont {{Margala}},
  \citenamefont {{Martini}}, \citenamefont {{Maus}}, \citenamefont {{McDonald}}, \citenamefont {{Medina-Varela}}, \citenamefont {{Meisner}}, \citenamefont {{Mena-Fern{\'a}ndez}}, \citenamefont {{Miquel}}, \citenamefont {{Moon}}, \citenamefont {{Moore}}, \citenamefont {{Moustakas}}, \citenamefont {{Mueller}}, \citenamefont {{Mu{\~n}oz-Guti{\'e}rrez}}, \citenamefont {{Myers}}, \citenamefont {{Nadathur}}, \citenamefont {{Napolitano}}, \citenamefont {{Neveux}}, \citenamefont {{Newman}}, \citenamefont {{Nguyen}}, \citenamefont {{Nie}}, \citenamefont {{Niz}}, \citenamefont {{Noriega}}, \citenamefont {{Padmanabhan}}, \citenamefont {{Paillas}}, \citenamefont {{Palanque-Delabrouille}}, \citenamefont {{Pan}}, \citenamefont {{Penmetsa}}, \citenamefont {{Percival}}, \citenamefont {{Pieri}}, \citenamefont {{Pinon}}, \citenamefont {{Poppett}}, \citenamefont {{Porredon}}, \citenamefont {{Prada}}, \citenamefont {{P{\'e}rez-Fern{\'a}ndez}}, \citenamefont {{P{\'e}rez-R{\`a}fols}}, \citenamefont {{Rabinowitz}}, \citenamefont
  {{Raichoor}}, \citenamefont {{Ram{\'\i}rez-P{\'e}rez}}, \citenamefont {{Ramirez-Solano}}, \citenamefont {{Rashkovetskyi}}, \citenamefont {{Ravoux}}, \citenamefont {{Rezaie}}, \citenamefont {{Rich}}, \citenamefont {{Rocher}}, \citenamefont {{Rockosi}}, \citenamefont {{Roe}}, \citenamefont {{Rosado-Marin}}, \citenamefont {{Ross}}, \citenamefont {{Rossi}}, \citenamefont {{Ruggeri}}, \citenamefont {{Ruhlmann-Kleider}}, \citenamefont {{Samushia}}, \citenamefont {{Sanchez}}, \citenamefont {{Saulder}}, \citenamefont {{Schlafly}}, \citenamefont {{Schlegel}}, \citenamefont {{Schubnell}}, \citenamefont {{Seo}}, \citenamefont {{Shafieloo}}, \citenamefont {{Sharples}}, \citenamefont {{Silber}}, \citenamefont {{Slosar}}, \citenamefont {{Smith}}, \citenamefont {{Sprayberry}}, \citenamefont {{Tan}}, \citenamefont {{Tarl{\'e}}}, \citenamefont {{Taylor}}, \citenamefont {{Trusov}}, \citenamefont {{Ure{\~n}a-L{\'o}pez}}, \citenamefont {{Vaisakh}}, \citenamefont {{Valcin}}, \citenamefont {{Valdes}}, \citenamefont
  {{Vargas-Maga{\~n}a}}, \citenamefont {{Verde}}, \citenamefont {{Walther}}, \citenamefont {{Wang}}, \citenamefont {{Wang}}, \citenamefont {{Weaver}}, \citenamefont {{Weaverdyck}}, \citenamefont {{Wechsler}}, \citenamefont {{Weinberg}}, \citenamefont {{White}}, \citenamefont {{Yu}}, \citenamefont {{Yu}}, \citenamefont {{Yuan}}, \citenamefont {{Y{\`e}che}}, \citenamefont {{Zaborowski}}, \citenamefont {{Zarrouk}}, \citenamefont {{Zhang}}, \citenamefont {{Zhao}}, \citenamefont {{Zhao}}, \citenamefont {{Zhou}},\ and\ \citenamefont {{Zhuang}}}]{DESI24BAOcosmo}%
  \BibitemOpen
  \bibfield  {author} {\bibinfo {author} {\bibfnamefont {A.~G.}\ \bibnamefont {{Adame}}}, \bibnamefont {et~al.},\ }\href {https://doi.org/10.1088/1475-7516/2025/02/021} {\bibfield  {journal} {\bibinfo  {journal} {\jcap}\ }\textbf {\bibinfo {volume} {2025}},\ \bibinfo {eid} {021} (\bibinfo {year} {2025})}\BibitemShut {NoStop}%
\bibitem [{\citenamefont {{Abdul Karim}}\ \emph {et~al.}(2025)\citenamefont {{Abdul Karim}}, \citenamefont {{Aguilar}}, \citenamefont {{Ahlen}}, \citenamefont {{Alam}}, \citenamefont {{Allen}}, \citenamefont {{Prieto}}, \citenamefont {{Alves}}, \citenamefont {{Anand}}, \citenamefont {{Andrade}}, \citenamefont {{Armengaud}}, \citenamefont {{Aviles}}, \citenamefont {{Bailey}}, \citenamefont {{Baltay}}, \citenamefont {{Bansal}}, \citenamefont {{Bault}}, \citenamefont {{Behera}}, \citenamefont {{BenZvi}}, \citenamefont {{Bianchi}}, \citenamefont {{Blake}}, \citenamefont {{Brieden}}, \citenamefont {{Brodzeller}}, \citenamefont {{Brooks}}, \citenamefont {{Buckley-Geer}}, \citenamefont {{Burtin}}, \citenamefont {{Calderon}}, \citenamefont {{Canning}}, \citenamefont {{Rosell}}, \citenamefont {{Carrilho}}, \citenamefont {{Casas}}, \citenamefont {{Castander}}, \citenamefont {{Charles}}, \citenamefont {{Chaussidon}}, \citenamefont {{Chaves-Montero}}, \citenamefont {{Chebat}}, \citenamefont {{Chen}}, \citenamefont
  {{Claybaugh}}, \citenamefont {{Cole}}, \citenamefont {{Cooper}}, \citenamefont {{Cuceu}}, \citenamefont {{Dawson}}, \citenamefont {{de la Macorra}}, \citenamefont {{de Mattia}}, \citenamefont {{Deiosso}}, \citenamefont {{Della Costa}}, \citenamefont {{Demina}}, \citenamefont {{Dey}}, \citenamefont {{Dey}}, \citenamefont {{Ding}}, \citenamefont {{Doel}}, \citenamefont {{Edelstein}}, \citenamefont {{Eisenstein}}, \citenamefont {{Elbers}}, \citenamefont {{Fagrelius}}, \citenamefont {{Fanning}}, \citenamefont {{Fern{\'a}ndez-Garc{\'\i}a}}, \citenamefont {{Ferraro}}, \citenamefont {{Font-Ribera}}, \citenamefont {{Forero-Romero}}, \citenamefont {{Frenk}}, \citenamefont {{Garcia-Quintero}}, \citenamefont {{Garrison}}, \citenamefont {{Gazta{\~n}aga}}, \citenamefont {{Gil-Mar{\'\i}n}}, \citenamefont {{Gontcho A Gontcho}}, \citenamefont {{Gonzalez}}, \citenamefont {{Gonzalez-Morales}}, \citenamefont {{Gordon}}, \citenamefont {{Green}}, \citenamefont {{Gutierrez}}, \citenamefont {{Guy}}, \citenamefont {{Hadzhiyska}},
  \citenamefont {{Hahn}}, \citenamefont {{He}}, \citenamefont {{Herbold}}, \citenamefont {{Herrera-Alcantar}}, \citenamefont {{Ho}}, \citenamefont {{Honscheid}}, \citenamefont {{Howlett}}, \citenamefont {{Huterer}}, \citenamefont {{Ishak}}, \citenamefont {{Juneau}}, \citenamefont {{Kamble}}, \citenamefont {{Kara{\c{c}}ayl{\i}}}, \citenamefont {{Kehoe}}, \citenamefont {{Kent}}, \citenamefont {{Kim}}, \citenamefont {{Kirkby}}, \citenamefont {{Kisner}}, \citenamefont {{Koposov}}, \citenamefont {{Kremin}}, \citenamefont {{Krolewski}}, \citenamefont {{Lahav}}, \citenamefont {{Lamman}}, \citenamefont {{Landriau}}, \citenamefont {{Lang}}, \citenamefont {{Lasker}}, \citenamefont {{Le Goff}}, \citenamefont {{Le Guillou}}, \citenamefont {{Leauthaud}}, \citenamefont {{Levi}}, \citenamefont {{Li}}, \citenamefont {{Li}}, \citenamefont {{Lodha}}, \citenamefont {{Lokken}}, \citenamefont {{Lozano-Rodr{\'\i}guez}}, \citenamefont {{Magneville}}, \citenamefont {{Manera}}, \citenamefont {{Martini}}, \citenamefont {{Matthewson}},
  \citenamefont {{Meisner}}, \citenamefont {{Mena-Fern{\'a}ndez}}, \citenamefont {{Menegas}}, \citenamefont {{Mergulh{\~a}o}}, \citenamefont {{Miquel}}, \citenamefont {{Moustakas}}, \citenamefont {{Mu{\~n}oz-Guti{\'e}rrez}}, \citenamefont {{Mu{\~n}oz-Santos}}, \citenamefont {{Myers}}, \citenamefont {{Nadathur}}, \citenamefont {{Naidoo}}, \citenamefont {{Napolitano}}, \citenamefont {{Newman}}, \citenamefont {{Niz}}, \citenamefont {{Noriega}}, \citenamefont {{Paillas}}, \citenamefont {{Palanque-Delabrouille}}, \citenamefont {{Pan}}, \citenamefont {{Peacock}}, \citenamefont {{Pellejero Ibanez}}, \citenamefont {{Percival}}, \citenamefont {{P{\'e}rez-Fern{\'a}ndez}}, \citenamefont {{P{\'e}rez-R{\`a}fols}}, \citenamefont {{Pieri}}, \citenamefont {{Poppett}}, \citenamefont {{Prada}}, \citenamefont {{Rabinowitz}}, \citenamefont {{Raichoor}}, \citenamefont {{Ram{\'\i}rez-P{\'e}rez}}, \citenamefont {{Rashkovetskyi}}, \citenamefont {{Ravoux}}, \citenamefont {{Rich}}, \citenamefont {{Rocher}}, \citenamefont {{Rockosi}},
  \citenamefont {{Rohlf}}, \citenamefont {{Rom{\'a}n-Herrera}}, \citenamefont {{Ross}}, \citenamefont {{Rossi}}, \citenamefont {{Ruggeri}}, \citenamefont {{Ruhlmann-Kleider}}, \citenamefont {{Samushia}}, \citenamefont {{Sanchez}}, \citenamefont {{Sanders}}, \citenamefont {{Schlegel}}, \citenamefont {{Schubnell}}, \citenamefont {{Seo}}, \citenamefont {{Shafieloo}}, \citenamefont {{Sharples}}, \citenamefont {{Silber}}, \citenamefont {{Sinigaglia}}, \citenamefont {{Sprayberry}}, \citenamefont {{Tan}}, \citenamefont {{Tarl{\'e}}}, \citenamefont {{Taylor}}, \citenamefont {{Turner}}, \citenamefont {{Ure{\~n}a-L{\'o}pez}}, \citenamefont {{Vaisakh}}, \citenamefont {{Valdes}}, \citenamefont {{Valogiannis}}, \citenamefont {{Vargas-Maga{\~n}a}}, \citenamefont {{Verde}}, \citenamefont {{Walther}}, \citenamefont {{Weaver}}, \citenamefont {{Weinberg}}, \citenamefont {{White}}, \citenamefont {{Wolfson}}, \citenamefont {{Y{\`e}che}}, \citenamefont {{Yu}}, \citenamefont {{Zaborowski}}, \citenamefont {{Zarrouk}}, \citenamefont
  {{Zhai}}, \citenamefont {{Zhang}}, \citenamefont {{Zhao}}, \citenamefont {{Zhao}}, \citenamefont {{Zhou}}, \citenamefont {{Zou}},\ and\ \citenamefont {{DESI Collaboration}}}]{DESI25BAOcosmo}%
  \BibitemOpen
  \bibfield  {author} {\bibinfo {author} {\bibfnamefont {M.}~\bibnamefont {{Abdul Karim}}}, \bibnamefont {et~al.},\ }\href {https://doi.org/10.1103/tr6y-kpc6} {\bibfield  {journal} {\bibinfo  {journal} {\prd}\ }\textbf {\bibinfo {volume} {112}},\ \bibinfo {eid} {083515} (\bibinfo {year} {2025})}\BibitemShut {NoStop}%
\bibitem [{\citenamefont {{Brout}}\ \emph {et~al.}(2022)\citenamefont {{Brout}}, \citenamefont {{Scolnic}}, \citenamefont {{Popovic}}, \citenamefont {{Riess}}, \citenamefont {{Carr}}, \citenamefont {{Zuntz}}, \citenamefont {{Kessler}}, \citenamefont {{Davis}}, \citenamefont {{Hinton}}, \citenamefont {{Jones}}, \citenamefont {{Kenworthy}}, \citenamefont {{Peterson}}, \citenamefont {{Said}}, \citenamefont {{Taylor}}, \citenamefont {{Ali}}, \citenamefont {{Armstrong}}, \citenamefont {{Charvu}}, \citenamefont {{Dwomoh}}, \citenamefont {{Meldorf}}, \citenamefont {{Palmese}}, \citenamefont {{Qu}}, \citenamefont {{Rose}}, \citenamefont {{Sanchez}}, \citenamefont {{Stubbs}}, \citenamefont {{Vincenzi}}, \citenamefont {{Wood}}, \citenamefont {{Brown}}, \citenamefont {{Chen}}, \citenamefont {{Chambers}}, \citenamefont {{Coulter}}, \citenamefont {{Dai}}, \citenamefont {{Dimitriadis}}, \citenamefont {{Filippenko}}, \citenamefont {{Foley}}, \citenamefont {{Jha}}, \citenamefont {{Kelsey}}, \citenamefont {{Kirshner}},
  \citenamefont {{M{\"o}ller}}, \citenamefont {{Muir}}, \citenamefont {{Nadathur}}, \citenamefont {{Pan}}, \citenamefont {{Rest}}, \citenamefont {{Rojas-Bravo}}, \citenamefont {{Sako}}, \citenamefont {{Siebert}}, \citenamefont {{Smith}}, \citenamefont {{Stahl}},\ and\ \citenamefont {{Wiseman}}}]{Brout22Pantheon+}%
  \BibitemOpen
  \bibfield  {author} {\bibinfo {author} {\bibfnamefont {D.}~\bibnamefont {{Brout}}}, \bibnamefont {et~al.},\ }\href {https://doi.org/10.3847/1538-4357/ac8e04} {\bibfield  {journal} {\bibinfo  {journal} {\apj}\ }\textbf {\bibinfo {volume} {938}},\ \bibinfo {eid} {110} (\bibinfo {year} {2022})}\BibitemShut {NoStop}%
\bibitem [{\citenamefont {{DES Collaboration}}\ \emph {et~al.}(2025)\citenamefont {{DES Collaboration}}, \citenamefont {{Abbott}}, \citenamefont {{Acevedo}}, \citenamefont {{Adamow}}, \citenamefont {{Aguena}}, \citenamefont {{Alarcon}}, \citenamefont {{Allam}}, \citenamefont {{Alves}}, \citenamefont {{Andrade-Oliveira}}, \citenamefont {{Annis}}, \citenamefont {{Armstrong}}, \citenamefont {{Avila}}, \citenamefont {{Bacon}}, \citenamefont {{Bechtol}}, \citenamefont {{Blazek}}, \citenamefont {{Bocquet}}, \citenamefont {{Brooks}}, \citenamefont {{Brout}}, \citenamefont {{Burke}}, \citenamefont {{Camacho}}, \citenamefont {{Camilleri}}, \citenamefont {{Campailla}}, \citenamefont {{Carnero Rosell}}, \citenamefont {{Carr}}, \citenamefont {{Carretero}}, \citenamefont {{Castander}}, \citenamefont {{Cawthon}}, \citenamefont {{Chan}}, \citenamefont {{Chang}}, \citenamefont {{Chen}}, \citenamefont {{Conselice}}, \citenamefont {{Costanzi}}, \citenamefont {{Crocce}} \emph {et~al.}}]{DES25BAO+SN}%
  \BibitemOpen
  \bibfield  {author} {\bibinfo {author} {\bibnamefont {{DES Collaboration}}}, \bibnamefont {et~al.},\ }\href {https://doi.org/10.48550/arXiv.2503.06712} {\bibfield  {journal} {\bibinfo  {journal} {arXiv e-prints}\ ,\ \bibinfo {eid} {arXiv:2503.06712}} (\bibinfo {year} {2025})}\BibitemShut {NoStop}%
\bibitem [{\citenamefont {{Salcedo}}\ \emph {et~al.}(2025{\natexlab{a}})\citenamefont {{Salcedo}}, \citenamefont {{Rozo}}, \citenamefont {{Wu}}, \citenamefont {{Cao}}, \citenamefont {{Paillas}}, \citenamefont {{Zhang}},\ and\ \citenamefont {{Rykoff}}}]{Salcedo2025b}%
  \BibitemOpen
  \bibfield  {author} {\bibinfo {author} {\bibfnamefont {A.~N.}\ \bibnamefont {{Salcedo}}}, \bibinfo {author} {\bibfnamefont {E.}~\bibnamefont {{Rozo}}}, \bibinfo {author} {\bibfnamefont {H.-Y.}\ \bibnamefont {{Wu}}}, \bibinfo {author} {\bibfnamefont {S.}~\bibnamefont {{Cao}}}, \bibinfo {author} {\bibfnamefont {E.}~\bibnamefont {{Paillas}}}, \bibinfo {author} {\bibfnamefont {H.}~\bibnamefont {{Zhang}}},\ \bibnamefont {and}\ \bibinfo {author} {\bibfnamefont {E.~S.}\ \bibnamefont {{Rykoff}}},\ }\href@noop {} {\bibfield  {journal} {\bibinfo  {journal} {arXiv e-prints}\ ,\ \bibinfo {eid} {arXiv:2512.06701}} (\bibinfo {year} {2025}{\natexlab{a}})}\BibitemShut {NoStop}%
\bibitem [{\citenamefont {{Planck Collaboration}}\ \emph {et~al.}(2020)\citenamefont {{Planck Collaboration}}, \citenamefont {{Aghanim}}, \citenamefont {{Akrami}}, \citenamefont {{Ashdown}}, \citenamefont {{Aumont}}, \citenamefont {{Baccigalupi}}, \citenamefont {{Ballardini}}, \citenamefont {{Banday}}, \citenamefont {{Barreiro}}, \citenamefont {{Bartolo}}, \citenamefont {{Basak}}, \citenamefont {{Battye}}, \citenamefont {{Benabed}}, \citenamefont {{Bernard}}, \citenamefont {{Bersanelli}}, \citenamefont {{Bielewicz}}, \citenamefont {{Bock}}, \citenamefont {{Bond}}, \citenamefont {{Borrill}}, \citenamefont {{Bouchet}}, \citenamefont {{Boulanger}}, \citenamefont {{Bucher}}, \citenamefont {{Burigana}}, \citenamefont {{Butler}}, \citenamefont {{Calabrese}}, \citenamefont {{Cardoso}}, \citenamefont {{Carron}}, \citenamefont {{Challinor}}, \citenamefont {{Chiang}}, \citenamefont {{Chluba}}, \citenamefont {{Colombo}}, \citenamefont {{Combet}}, \citenamefont {{Contreras}}, \citenamefont {{Crill}}, \citenamefont
  {{Cuttaia}}, \citenamefont {{de Bernardis}}, \citenamefont {{de Zotti}}, \citenamefont {{Delabrouille}}, \citenamefont {{Delouis}}, \citenamefont {{Di Valentino}}, \citenamefont {{Diego}}, \citenamefont {{Dor{\'e}}}, \citenamefont {{Douspis}}, \citenamefont {{Ducout}}, \citenamefont {{Dupac}}, \citenamefont {{Dusini}}, \citenamefont {{Efstathiou}}, \citenamefont {{Elsner}}, \citenamefont {{En{\ss}lin}}, \citenamefont {{Eriksen}}, \citenamefont {{Fantaye}}, \citenamefont {{Farhang}}, \citenamefont {{Fergusson}}, \citenamefont {{Fernandez-Cobos}}, \citenamefont {{Finelli}}, \citenamefont {{Forastieri}}, \citenamefont {{Frailis}}, \citenamefont {{Fraisse}}, \citenamefont {{Franceschi}}, \citenamefont {{Frolov}}, \citenamefont {{Galeotta}}, \citenamefont {{Galli}}, \citenamefont {{Ganga}}, \citenamefont {{G{\'e}nova-Santos}}, \citenamefont {{Gerbino}}, \citenamefont {{Ghosh}}, \citenamefont {{Gonz{\'a}lez-Nuevo}}, \citenamefont {{G{\'o}rski}}, \citenamefont {{Gratton}}, \citenamefont {{Gruppuso}}, \citenamefont
  {{Gudmundsson}}, \citenamefont {{Hamann}}, \citenamefont {{Handley}}, \citenamefont {{Hansen}}, \citenamefont {{Herranz}}, \citenamefont {{Hildebrandt}}, \citenamefont {{Hivon}}, \citenamefont {{Huang}}, \citenamefont {{Jaffe}}, \citenamefont {{Jones}}, \citenamefont {{Karakci}}, \citenamefont {{Keih{\"a}nen}}, \citenamefont {{Keskitalo}}, \citenamefont {{Kiiveri}}, \citenamefont {{Kim}}, \citenamefont {{Kisner}}, \citenamefont {{Knox}}, \citenamefont {{Krachmalnicoff}}, \citenamefont {{Kunz}}, \citenamefont {{Kurki-Suonio}}, \citenamefont {{Lagache}}, \citenamefont {{Lamarre}}, \citenamefont {{Lasenby}}, \citenamefont {{Lattanzi}}, \citenamefont {{Lawrence}}, \citenamefont {{Le Jeune}}, \citenamefont {{Lemos}}, \citenamefont {{Lesgourgues}}, \citenamefont {{Levrier}}, \citenamefont {{Lewis}}, \citenamefont {{Liguori}}, \citenamefont {{Lilje}}, \citenamefont {{Lilley}}, \citenamefont {{Lindholm}}, \citenamefont {{L{\'o}pez-Caniego}}, \citenamefont {{Lubin}}, \citenamefont {{Ma}}, \citenamefont
  {{Mac{\'\i}as-P{\'e}rez}}, \citenamefont {{Maggio}}, \citenamefont {{Maino}}, \citenamefont {{Mandolesi}}, \citenamefont {{Mangilli}}, \citenamefont {{Marcos-Caballero}}, \citenamefont {{Maris}}, \citenamefont {{Martin}}, \citenamefont {{Martinelli}}, \citenamefont {{Mart{\'\i}nez-Gonz{\'a}lez}}, \citenamefont {{Matarrese}}, \citenamefont {{Mauri}}, \citenamefont {{McEwen}}, \citenamefont {{Meinhold}}, \citenamefont {{Melchiorri}}, \citenamefont {{Mennella}}, \citenamefont {{Migliaccio}}, \citenamefont {{Millea}}, \citenamefont {{Mitra}}, \citenamefont {{Miville-Desch{\^e}nes}}, \citenamefont {{Molinari}}, \citenamefont {{Montier}}, \citenamefont {{Morgante}}, \citenamefont {{Moss}}, \citenamefont {{Natoli}}, \citenamefont {{N{\o}rgaard-Nielsen}}, \citenamefont {{Pagano}}, \citenamefont {{Paoletti}}, \citenamefont {{Partridge}}, \citenamefont {{Patanchon}}, \citenamefont {{Peiris}}, \citenamefont {{Perrotta}}, \citenamefont {{Pettorino}}, \citenamefont {{Piacentini}}, \citenamefont {{Polastri}},
  \citenamefont {{Polenta}}, \citenamefont {{Puget}}, \citenamefont {{Rachen}}, \citenamefont {{Reinecke}}, \citenamefont {{Remazeilles}}, \citenamefont {{Renzi}}, \citenamefont {{Rocha}}, \citenamefont {{Rosset}}, \citenamefont {{Roudier}}, \citenamefont {{Rubi{\~n}o-Mart{\'\i}n}}, \citenamefont {{Ruiz-Granados}}, \citenamefont {{Salvati}}, \citenamefont {{Sandri}}, \citenamefont {{Savelainen}}, \citenamefont {{Scott}}, \citenamefont {{Shellard}}, \citenamefont {{Sirignano}}, \citenamefont {{Sirri}}, \citenamefont {{Spencer}}, \citenamefont {{Sunyaev}}, \citenamefont {{Suur-Uski}}, \citenamefont {{Tauber}}, \citenamefont {{Tavagnacco}}, \citenamefont {{Tenti}}, \citenamefont {{Toffolatti}}, \citenamefont {{Tomasi}}, \citenamefont {{Trombetti}}, \citenamefont {{Valenziano}}, \citenamefont {{Valiviita}}, \citenamefont {{Van Tent}}, \citenamefont {{Vibert}}, \citenamefont {{Vielva}}, \citenamefont {{Villa}}, \citenamefont {{Vittorio}}, \citenamefont {{Wandelt}}, \citenamefont {{Wehus}}, \citenamefont {{White}},
  \citenamefont {{White}}, \citenamefont {{Zacchei}},\ and\ \citenamefont {{Zonca}}}]{Planck18cosmo}%
  \BibitemOpen
  \bibfield  {author} {\bibinfo {author} {\bibnamefont {{Planck Collaboration}}}, \bibnamefont {et~al.},\ }\href {https://doi.org/10.1051/0004-6361/201833910} {\bibfield  {journal} {\bibinfo  {journal} {\aap}\ }\textbf {\bibinfo {volume} {641}},\ \bibinfo {eid} {A6} (\bibinfo {year} {2020})}\BibitemShut {NoStop}%
\bibitem [{\citenamefont {{Li}}\ \emph {et~al.}(2022)\citenamefont {{Li}}, \citenamefont {{Miyatake}}, \citenamefont {{Luo}}, \citenamefont {{More}}, \citenamefont {{Oguri}}, \citenamefont {{Hamana}}, \citenamefont {{Mandelbaum}}, \citenamefont {{Shirasaki}}, \citenamefont {{Takada}}, \citenamefont {{Armstrong}}, \citenamefont {{Kannawadi}}, \citenamefont {{Takita}}, \citenamefont {{Miyazaki}}, \citenamefont {{Nishizawa}}, \citenamefont {{Plazas Malagon}}, \citenamefont {{Strauss}}, \citenamefont {{Tanaka}},\ and\ \citenamefont {{Yoshida}}}]{Li22HSC}%
  \BibitemOpen
  \bibfield  {author} {\bibinfo {author} {\bibfnamefont {X.}~\bibnamefont {{Li}}}, \bibnamefont {et~al.},\ }\href {https://doi.org/10.1093/pasj/psac006} {\bibfield  {journal} {\bibinfo  {journal} {\pasj}\ }\textbf {\bibinfo {volume} {74}},\ \bibinfo {pages} {421} (\bibinfo {year} {2022})}\BibitemShut {NoStop}%
\bibitem [{\citenamefont {{Dalal}}\ \emph {et~al.}(2023)\citenamefont {{Dalal}}, \citenamefont {{Li}}, \citenamefont {{Nicola}}, \citenamefont {{Zuntz}}, \citenamefont {{Strauss}}, \citenamefont {{Sugiyama}}, \citenamefont {{Zhang}}, \citenamefont {{Rau}}, \citenamefont {{Mandelbaum}}, \citenamefont {{Takada}}, \citenamefont {{More}}, \citenamefont {{Miyatake}}, \citenamefont {{Kannawadi}}, \citenamefont {{Shirasaki}}, \citenamefont {{Taniguchi}}, \citenamefont {{Takahashi}}, \citenamefont {{Osato}}, \citenamefont {{Hamana}}, \citenamefont {{Oguri}}, \citenamefont {{Nishizawa}}, \citenamefont {{Malag{\'o}n}}, \citenamefont {{Sunayama}}, \citenamefont {{Alonso}}, \citenamefont {{Slosar}}, \citenamefont {{Luo}}, \citenamefont {{Armstrong}}, \citenamefont {{Bosch}}, \citenamefont {{Hsieh}}, \citenamefont {{Komiyama}}, \citenamefont {{Lupton}}, \citenamefont {{Lust}}, \citenamefont {{MacArthur}}, \citenamefont {{Miyazaki}}, \citenamefont {{Murayama}}, \citenamefont {{Nishimichi}}, \citenamefont {{Okura}},
  \citenamefont {{Price}}, \citenamefont {{Tait}}, \citenamefont {{Tanaka}},\ and\ \citenamefont {{Wang}}}]{Dalal23HSC}%
  \BibitemOpen
  \bibfield  {author} {\bibinfo {author} {\bibfnamefont {R.}~\bibnamefont {{Dalal}}}, \bibnamefont {et~al.},\ }\href {https://doi.org/10.1103/PhysRevD.108.123519} {\bibfield  {journal} {\bibinfo  {journal} {\prd}\ }\textbf {\bibinfo {volume} {108}},\ \bibinfo {eid} {123519} (\bibinfo {year} {2023})}\BibitemShut {NoStop}%
\bibitem [{\citenamefont {{Amon}}\ \emph {et~al.}(2023)\citenamefont {{Amon}}, \citenamefont {{Robertson}}, \citenamefont {{Miyatake}}, \citenamefont {{Heymans}}, \citenamefont {{White}}, \citenamefont {{DeRose}}, \citenamefont {{Yuan}}, \citenamefont {{Wechsler}}, \citenamefont {{Varga}}, \citenamefont {{Bocquet}}, \citenamefont {{Dvornik}}, \citenamefont {{More}}, \citenamefont {{Ross}}, \citenamefont {{Hoekstra}}, \citenamefont {{Alarcon}}, \citenamefont {{Asgari}}, \citenamefont {{Blazek}}, \citenamefont {{Campos}}, \citenamefont {{Chen}}, \citenamefont {{Choi}}, \citenamefont {{Crocce}} \emph {et~al.}}]{AmonRobertson23}%
  \BibitemOpen
  \bibfield  {author} {\bibinfo {author} {\bibfnamefont {A.}~\bibnamefont {{Amon}}}, \bibnamefont {et~al.},\ }\href {https://doi.org/10.1093/mnras/stac2938} {\bibfield  {journal} {\bibinfo  {journal} {\mnras}\ }\textbf {\bibinfo {volume} {518}},\ \bibinfo {pages} {477} (\bibinfo {year} {2023})}\BibitemShut {NoStop}%
\bibitem [{\citenamefont {{Eifler}}\ \emph {et~al.}(2021)\citenamefont {{Eifler}}, \citenamefont {{Miyatake}}, \citenamefont {{Krause}}, \citenamefont {{Heinrich}}, \citenamefont {{Miranda}}, \citenamefont {{Hirata}}, \citenamefont {{Xu}}, \citenamefont {{Hemmati}}, \citenamefont {{Simet}}, \citenamefont {{Capak}}, \citenamefont {{Choi}}, \citenamefont {{Dor{\'e}}}, \citenamefont {{Doux}}, \citenamefont {{Fang}}, \citenamefont {{Hounsell}}, \citenamefont {{Huff}}, \citenamefont {{Huang}}, \citenamefont {{Jarvis}}, \citenamefont {{Kruk}}, \citenamefont {{Masters}}, \citenamefont {{Rozo}}, \citenamefont {{Scolnic}}, \citenamefont {{Spergel}}, \citenamefont {{Troxel}}, \citenamefont {{von der Linden}}, \citenamefont {{Wang}}, \citenamefont {{Weinberg}}, \citenamefont {{Wenzl}},\ and\ \citenamefont {{Wu}}}]{Eifler21}%
  \BibitemOpen
  \bibfield  {author} {\bibinfo {author} {\bibfnamefont {T.}~\bibnamefont {{Eifler}}}, \bibnamefont {et~al.},\ }\href {https://doi.org/10.1093/mnras/stab1762} {\bibfield  {journal} {\bibinfo  {journal} {\mnras}\ }\textbf {\bibinfo {volume} {507}},\ \bibinfo {pages} {1746} (\bibinfo {year} {2021})}\BibitemShut {NoStop}%
\bibitem [{\citenamefont {{To}}\ \emph {et~al.}(2021)\citenamefont {{To}}, \citenamefont {{Krause}}, \citenamefont {{Rozo}}, \citenamefont {{Wu}}, \citenamefont {{Gruen}}, \citenamefont {{Wechsler}}, \citenamefont {{Eifler}}, \citenamefont {{Rykoff}}, \citenamefont {{Costanzi}}, \citenamefont {{Becker}}, \citenamefont {{Bernstein}}, \citenamefont {{Blazek}}, \citenamefont {{Bocquet}}, \citenamefont {{Bridle}}, \citenamefont {{Cawthon}}, \citenamefont {{Choi}}, \citenamefont {{Crocce}}, \citenamefont {{Davis}}, \citenamefont {{DeRose}}, \citenamefont {{Drlica-Wagner}}, \citenamefont {{Elvin-Poole}}, \citenamefont {{Fang}}, \citenamefont {{Farahi}}, \citenamefont {{Friedrich}}, \citenamefont {{Gatti}}, \citenamefont {{Gaztanaga}}, \citenamefont {{Giannantonio}}, \citenamefont {{Hartley}}, \citenamefont {{Hoyle}}, \citenamefont {{Jarvis}}, \citenamefont {{MacCrann}}, \citenamefont {{McClintock}}, \citenamefont {{Miranda}}, \citenamefont {{Pereira}}, \citenamefont {{Park}}, \citenamefont {{Porredon}},
  \citenamefont {{Prat}}, \citenamefont {{Rau}}, \citenamefont {{Ross}}, \citenamefont {{Samuroff}}, \citenamefont {{S{\'a}nchez}}, \citenamefont {{Sevilla-Noarbe}}, \citenamefont {{Sheldon}}, \citenamefont {{Troxel}}, \citenamefont {{Varga}}, \citenamefont {{Vielzeuf}}, \citenamefont {{Zhang}}, \citenamefont {{Zuntz}}, \citenamefont {{Abbott}}, \citenamefont {{Aguena}}, \citenamefont {{Amon}}, \citenamefont {{Annis}}, \citenamefont {{Avila}}, \citenamefont {{Bertin}}, \citenamefont {{Bhargava}}, \citenamefont {{Brooks}}, \citenamefont {{Burke}}, \citenamefont {{Carnero Rosell}}, \citenamefont {{Carrasco Kind}}, \citenamefont {{Carretero}}, \citenamefont {{Chang}}, \citenamefont {{Conselice}}, \citenamefont {{da Costa}}, \citenamefont {{Davis}}, \citenamefont {{Desai}}, \citenamefont {{Diehl}}, \citenamefont {{Dietrich}}, \citenamefont {{Everett}}, \citenamefont {{Evrard}}, \citenamefont {{Ferrero}}, \citenamefont {{Flaugher}}, \citenamefont {{Fosalba}}, \citenamefont {{Frieman}}, \citenamefont
  {{Garc{\'\i}a-Bellido}}, \citenamefont {{Gruendl}}, \citenamefont {{Gutierrez}}, \citenamefont {{Hinton}}, \citenamefont {{Hollowood}}, \citenamefont {{Honscheid}}, \citenamefont {{Huterer}}, \citenamefont {{James}}, \citenamefont {{Jeltema}}, \citenamefont {{Kron}}, \citenamefont {{Kuehn}}, \citenamefont {{Kuropatkin}}, \citenamefont {{Lima}}, \citenamefont {{Maia}}, \citenamefont {{Marshall}}, \citenamefont {{Menanteau}}, \citenamefont {{Miquel}}, \citenamefont {{Morgan}}, \citenamefont {{Muir}}, \citenamefont {{Myles}}, \citenamefont {{Palmese}}, \citenamefont {{Paz-Chinch{\'o}n}}, \citenamefont {{Plazas}}, \citenamefont {{Romer}}, \citenamefont {{Roodman}}, \citenamefont {{Sanchez}}, \citenamefont {{Santiago}}, \citenamefont {{Scarpine}}, \citenamefont {{Serrano}}, \citenamefont {{Smith}}, \citenamefont {{Suchyta}}, \citenamefont {{Swanson}}, \citenamefont {{Tarle}}, \citenamefont {{Thomas}}, \citenamefont {{Tucker}}, \citenamefont {{Weller}}, \citenamefont {{Wester}}, \citenamefont {{Wilkinson}},\ and\
  \citenamefont {{DES Collaboration}}}]{DESY14x2ptN}%
  \BibitemOpen
  \bibfield  {author} {\bibinfo {author} {\bibfnamefont {C.-H.}\ \bibnamefont {{To}}}, \bibnamefont {et~al.},\ }\href {https://doi.org/10.1103/PhysRevLett.126.141301} {\bibfield  {journal} {\bibinfo  {journal} {\prl}\ }\textbf {\bibinfo {volume} {126}},\ \bibinfo {eid} {141301} (\bibinfo {year} {2021})}\BibitemShut {NoStop}%
\bibitem [{\citenamefont {{Abbott}}\ \emph {et~al.}(2025)\citenamefont {{Abbott}}, \citenamefont {{Aguena}}, \citenamefont {{Alarcon}}, \citenamefont {{Amon}}, \citenamefont {{Anbajagane}}, \citenamefont {{Andrade-Oliveira}}, \citenamefont {{Avila}}, \citenamefont {{Allam}}, \citenamefont {{Bacon}}, \citenamefont {{Becker}}, \citenamefont {{Bhargava}}, \citenamefont {{Blazek}}, \citenamefont {{Bocquet}}, \citenamefont {{Brooks}}, \citenamefont {{Rosell}}, \citenamefont {{Carretero}}, \citenamefont {{Castander}}, \citenamefont {{Chang}}, \citenamefont {{Choi}}, \citenamefont {{Conselice}}, \citenamefont {{Costanzi}}, \citenamefont {{Crocce}}, \citenamefont {{da Costa}}, \citenamefont {{Pereira}}, \citenamefont {{Doux}}, \citenamefont {{Davis}}, \citenamefont {{Desai}}, \citenamefont {{Diehl}}, \citenamefont {{Dodelson}}, \citenamefont {{Doel}}, \citenamefont {{Elvin-Poole}}, \citenamefont {{Esteves}}, \citenamefont {{Everett}}, \citenamefont {{Farahi}}, \citenamefont {{Fert{\'e}}}, \citenamefont {{Flaugher}},
  \citenamefont {{Garc{\'\i}a-Bellido}}, \citenamefont {{Gatti}}, \citenamefont {{Giannini}}, \citenamefont {{Giles}}, \citenamefont {{Grandis}}, \citenamefont {{Gruen}}, \citenamefont {{Gruendl}}, \citenamefont {{Gutierrez}}, \citenamefont {{Harrison}}, \citenamefont {{Hinton}}, \citenamefont {{Hollowood}}, \citenamefont {{Honscheid}}, \citenamefont {{Jeffrey}}, \citenamefont {{Jeltema}}, \citenamefont {{Krause}}, \citenamefont {{Lahav}}, \citenamefont {{Lee}}, \citenamefont {{Lidman}}, \citenamefont {{Lima}}, \citenamefont {{Lin}}, \citenamefont {{Mohr}}, \citenamefont {{Marshall}}, \citenamefont {{McCullough}}, \citenamefont {{Mena-Fern'andez}}, \citenamefont {{Miquel}}, \citenamefont {{Muir}}, \citenamefont {{Myles}}, \citenamefont {{Ogando}}, \citenamefont {{Palmese}}, \citenamefont {{Paterno}}, \citenamefont {{Malag{\'o}n}}, \citenamefont {{Porredon}}, \citenamefont {{Prat}}, \citenamefont {{Romer}}, \citenamefont {{Roodman}}, \citenamefont {{Rozo}}, \citenamefont {{Rykoff}}, \citenamefont
  {{Rosenfeld}}, \citenamefont {{Sanchez}}, \citenamefont {{Cid}}, \citenamefont {{Sevilla-Noarbe}}, \citenamefont {{Smith}}, \citenamefont {{Suchyta}}, \citenamefont {{Shin}}, \citenamefont {{Tarle}}, \citenamefont {{Thomas}}, \citenamefont {{To}}, \citenamefont {{Troxel}}, \citenamefont {{Vikram}}, \citenamefont {{Walker}}, \citenamefont {{Weinberg}}, \citenamefont {{Weaverdyck}}, \citenamefont {{Wechsler}}, \citenamefont {{Weller}}, \citenamefont {{Wu}}, \citenamefont {{Yamamoto}}, \citenamefont {{Yanny}}, \citenamefont {{Zhang}}, \citenamefont {{Zhou}},\ and\ \citenamefont {{DES Collaboration}}}]{DESY3CL+3x2pt}%
  \BibitemOpen
  \bibfield  {author} {\bibinfo {author} {\bibfnamefont {T.~M.~C.}\ \bibnamefont {{Abbott}}}, \bibnamefont {et~al.},\ }\href {https://doi.org/10.1103/3dzh-d8f5} {\bibfield  {journal} {\bibinfo  {journal} {\prd}\ }\textbf {\bibinfo {volume} {112}},\ \bibinfo {eid} {083535} (\bibinfo {year} {2025})}\BibitemShut {NoStop}%
\bibitem [{\citenamefont {{To}}\ \emph {et~al.}(2025)\citenamefont {{To}}, \citenamefont {{Krause}}, \citenamefont {{Chang}}, \citenamefont {{Wu}}, \citenamefont {{Wechsler}}, \citenamefont {{Rozo}}, \citenamefont {{Weinberg}}, \citenamefont {{Anbajagane}}, \citenamefont {{Avila}}, \citenamefont {{Blazek}}, \citenamefont {{Bocquet}}, \citenamefont {{Costanzi}}, \citenamefont {{De Vicente}}, \citenamefont {{Elvin-Poole}}, \citenamefont {{Fert{\'e}}}, \citenamefont {{Grandis}}, \citenamefont {{Muir}}, \citenamefont {{Porredon}}, \citenamefont {{Samuroff}}, \citenamefont {{Sanchez}}, \citenamefont {{Cid}}, \citenamefont {{Sevilla-Noarbe}}, \citenamefont {{Weaverdyck}}, \citenamefont {{Abbott}}, \citenamefont {{Aguena}}, \citenamefont {{Andrade-Oliveira}}, \citenamefont {{Bacon}}, \citenamefont {{Becker}}, \citenamefont {{Brooks}}, \citenamefont {{Rosell}}, \citenamefont {{Carretero}}, \citenamefont {{Choi}}, \citenamefont {{da Costa}}, \citenamefont {{Pereira}}, \citenamefont {{Davis}}, \citenamefont {{Desai}},
  \citenamefont {{Doel}}, \citenamefont {{Doux}}, \citenamefont {{Everett}}, \citenamefont {{Frieman}}, \citenamefont {{Garc{\'\i}a-Bellido}}, \citenamefont {{Gatti}}, \citenamefont {{Gaztanaga}}, \citenamefont {{Giannini}}, \citenamefont {{Gruen}}, \citenamefont {{Gutierrez}}, \citenamefont {{Hinton}}, \citenamefont {{Hollowood}}, \citenamefont {{Honscheid}}, \citenamefont {{Jeltema}}, \citenamefont {{Kuehn}}, \citenamefont {{Lee}}, \citenamefont {{Marshall}}, \citenamefont {{Mena-Fern{\'a}ndez}}, \citenamefont {{Miquel}}, \citenamefont {{Mohr}}, \citenamefont {{Myles}}, \citenamefont {{Palmese}}, \citenamefont {{Malag{\'o}n}}, \citenamefont {{Romer}}, \citenamefont {{Shin}}, \citenamefont {{Smith}}, \citenamefont {{Suchyta}}, \citenamefont {{Tarle}}, \citenamefont {{Vikram}}, \citenamefont {{Walker}}, \citenamefont {{Weller}},\ and\ \citenamefont {{DES Collaboration}}}]{To25}%
  \BibitemOpen
  \bibfield  {author} {\bibinfo {author} {\bibfnamefont {C.-H.}\ \bibnamefont {{To}}}, \bibnamefont {et~al.},\ }\href {https://doi.org/10.1103/ynqj-6hsb} {\bibfield  {journal} {\bibinfo  {journal} {\prd}\ }\textbf {\bibinfo {volume} {112}},\ \bibinfo {eid} {063537} (\bibinfo {year} {2025})}\BibitemShut {NoStop}%
\bibitem [{\citenamefont {{Lima}}\ and\ \citenamefont {{Hu}}(2005)}]{LimaHu2005}%
  \BibitemOpen
  \bibfield  {author} {\bibinfo {author} {\bibfnamefont {M.}~\bibnamefont {{Lima}}}\ \bibnamefont {and}\ \bibinfo {author} {\bibfnamefont {W.}~\bibnamefont {{Hu}}},\ }\href {https://doi.org/10.1103/PhysRevD.72.043006} {\bibfield  {journal} {\bibinfo  {journal} {\prd}\ }\textbf {\bibinfo {volume} {72}},\ \bibinfo {eid} {043006} (\bibinfo {year} {2005})}\BibitemShut {NoStop}%
\bibitem [{\citenamefont {{Abbott}}\ \emph {et~al.}(2020)\citenamefont {{Abbott}}, \citenamefont {{Aguena}}, \citenamefont {{Alarcon}}, \citenamefont {{Allam}}, \citenamefont {{Allen}}, \citenamefont {{Annis}}, \citenamefont {{Avila}}, \citenamefont {{Bacon}}, \citenamefont {{Bechtol}}, \citenamefont {{Bermeo}}, \citenamefont {{Bernstein}}, \citenamefont {{Bertin}}, \citenamefont {{Bhargava}}, \citenamefont {{Bocquet}}, \citenamefont {{Brooks}}, \citenamefont {{Brout}}, \citenamefont {{Buckley-Geer}}, \citenamefont {{Burke}}, \citenamefont {{Carnero Rosell}}, \citenamefont {{Carrasco Kind}}, \citenamefont {{Carretero}}, \citenamefont {{Castander}}, \citenamefont {{Cawthon}}, \citenamefont {{Chang}}, \citenamefont {{Chen}}, \citenamefont {{Choi}}, \citenamefont {{Costanzi}}, \citenamefont {{Crocce}}, \citenamefont {{da Costa}}, \citenamefont {{Davis}}, \citenamefont {{De Vicente}}, \citenamefont {{DeRose}}, \citenamefont {{Desai}}, \citenamefont {{Diehl}}, \citenamefont {{Dietrich}}, \citenamefont
  {{Dodelson}}, \citenamefont {{Doel}}, \citenamefont {{Drlica-Wagner}}, \citenamefont {{Eckert}}, \citenamefont {{Eifler}}, \citenamefont {{Elvin-Poole}}, \citenamefont {{Estrada}}, \citenamefont {{Everett}}, \citenamefont {{Evrard}}, \citenamefont {{Farahi}}, \citenamefont {{Ferrero}}, \citenamefont {{Flaugher}}, \citenamefont {{Fosalba}}, \citenamefont {{Frieman}}, \citenamefont {{Garc{\'\i}a-Bellido}}, \citenamefont {{Gatti}}, \citenamefont {{Gaztanaga}}, \citenamefont {{Gerdes}}, \citenamefont {{Giannantonio}}, \citenamefont {{Giles}}, \citenamefont {{Grandis}}, \citenamefont {{Gruen}}, \citenamefont {{Gruendl}}, \citenamefont {{Gschwend}}, \citenamefont {{Gutierrez}}, \citenamefont {{Hartley}}, \citenamefont {{Hinton}}, \citenamefont {{Hollowood}}, \citenamefont {{Honscheid}}, \citenamefont {{Hoyle}}, \citenamefont {{Huterer}}, \citenamefont {{James}}, \citenamefont {{Jarvis}}, \citenamefont {{Jeltema}}, \citenamefont {{Johnson}}, \citenamefont {{Johnson}}, \citenamefont {{Kent}}, \citenamefont
  {{Krause}}, \citenamefont {{Kron}}, \citenamefont {{Kuehn}}, \citenamefont {{Kuropatkin}}, \citenamefont {{Lahav}}, \citenamefont {{Li}}, \citenamefont {{Lidman}}, \citenamefont {{Lima}}, \citenamefont {{Lin}}, \citenamefont {{MacCrann}}, \citenamefont {{Maia}}, \citenamefont {{Mantz}}, \citenamefont {{Marshall}}, \citenamefont {{Martini}}, \citenamefont {{Mayers}}, \citenamefont {{Melchior}}, \citenamefont {{Mena-Fern{\'a}ndez}}, \citenamefont {{Menanteau}}, \citenamefont {{Miquel}}, \citenamefont {{Mohr}}, \citenamefont {{Nichol}}, \citenamefont {{Nord}}, \citenamefont {{Ogando}}, \citenamefont {{Palmese}}, \citenamefont {{Paz-Chinch{\'o}n}}, \citenamefont {{Plazas}}, \citenamefont {{Prat}}, \citenamefont {{Rau}}, \citenamefont {{Romer}}, \citenamefont {{Roodman}}, \citenamefont {{Rooney}}, \citenamefont {{Rozo}}, \citenamefont {{Rykoff}}, \citenamefont {{Sako}}, \citenamefont {{Samuroff}}, \citenamefont {{S{\'a}nchez}}, \citenamefont {{Sanchez}}, \citenamefont {{Saro}}, \citenamefont {{Scarpine}},
  \citenamefont {{Schubnell}}, \citenamefont {{Scolnic}}, \citenamefont {{Serrano}}, \citenamefont {{Sevilla-Noarbe}}, \citenamefont {{Sheldon}}, \citenamefont {{Smith}}, \citenamefont {{Smith}}, \citenamefont {{Suchyta}}, \citenamefont {{Swanson}}, \citenamefont {{Tarle}}, \citenamefont {{Thomas}}, \citenamefont {{To}}, \citenamefont {{Troxel}}, \citenamefont {{Tucker}}, \citenamefont {{Varga}}, \citenamefont {{von der Linden}}, \citenamefont {{Walker}}, \citenamefont {{Wechsler}}, \citenamefont {{Weller}}, \citenamefont {{Wilkinson}}, \citenamefont {{Wu}}, \citenamefont {{Yanny}}, \citenamefont {{Zhang}}, \citenamefont {{Zhang}}, \citenamefont {{Zuntz}},\ and\ \citenamefont {{DES Collaboration}}}]{DESY1CL}%
  \BibitemOpen
  \bibfield  {author} {\bibinfo {author} {\bibfnamefont {T.~M.~C.}\ \bibnamefont {{Abbott}}}, \bibnamefont {et~al.},\ }\href {https://doi.org/10.1103/PhysRevD.102.023509} {\bibfield  {journal} {\bibinfo  {journal} {\prd}\ }\textbf {\bibinfo {volume} {102}},\ \bibinfo {eid} {023509} (\bibinfo {year} {2020})}\BibitemShut {NoStop}%
\bibitem [{\citenamefont {{Salcedo}}\ \emph {et~al.}(2020)\citenamefont {{Salcedo}}, \citenamefont {{Wibking}}, \citenamefont {{Weinberg}}, \citenamefont {{Wu}}, \citenamefont {{Ferrer}}, \citenamefont {{Eisenstein}},\ and\ \citenamefont {{Pinto}}}]{Salcedoetal2020}%
  \BibitemOpen
  \bibfield  {author} {\bibinfo {author} {\bibfnamefont {A.~N.}\ \bibnamefont {{Salcedo}}}, \bibinfo {author} {\bibfnamefont {B.~D.}\ \bibnamefont {{Wibking}}}, \bibinfo {author} {\bibfnamefont {D.~H.}\ \bibnamefont {{Weinberg}}}, \bibinfo {author} {\bibfnamefont {H.-Y.}\ \bibnamefont {{Wu}}}, \bibinfo {author} {\bibfnamefont {D.}~\bibnamefont {{Ferrer}}}, \bibinfo {author} {\bibfnamefont {D.}~\bibnamefont {{Eisenstein}}},\ \bibnamefont {and}\ \bibinfo {author} {\bibfnamefont {P.}~\bibnamefont {{Pinto}}},\ }\href {https://doi.org/10.1093/mnras/stz2963} {\bibfield  {journal} {\bibinfo  {journal} {\mnras}\ }\textbf {\bibinfo {volume} {491}},\ \bibinfo {pages} {3061} (\bibinfo {year} {2020})}\BibitemShut {NoStop}%
\bibitem [{\citenamefont {{Wechsler}}\ and\ \citenamefont {{Tinker}}(2018)}]{WechslerTinker18}%
  \BibitemOpen
  \bibfield  {author} {\bibinfo {author} {\bibfnamefont {R.~H.}\ \bibnamefont {{Wechsler}}}\ \bibnamefont {and}\ \bibinfo {author} {\bibfnamefont {J.~L.}\ \bibnamefont {{Tinker}}},\ }\href {https://doi.org/10.1146/annurev-astro-081817-051756} {\bibfield  {journal} {\bibinfo  {journal} {\araa}\ }\textbf {\bibinfo {volume} {56}},\ \bibinfo {pages} {435} (\bibinfo {year} {2018})}\BibitemShut {NoStop}%
\bibitem [{\citenamefont {{Zacharegkas}}\ \emph {et~al.}(2022)\citenamefont {{Zacharegkas}}, \citenamefont {{Chang}}, \citenamefont {{Prat}}, \citenamefont {{Pandey}}, \citenamefont {{Ferrero}}, \citenamefont {{Blazek}}, \citenamefont {{Jain}}, \citenamefont {{Crocce}}, \citenamefont {{DeRose}}, \citenamefont {{Palmese}}, \citenamefont {{Seitz}}, \citenamefont {{Sheldon}}, \citenamefont {{Hartley}}, \citenamefont {{Wechsler}}, \citenamefont {{Dodelson}}, \citenamefont {{Fosalba}}, \citenamefont {{Krause}}, \citenamefont {{Park}}, \citenamefont {{S{\'a}nchez}}, \citenamefont {{Alarcon}}, \citenamefont {{Amon}}, \citenamefont {{Bechtol}}, \citenamefont {{Becker}}, \citenamefont {{Bernstein}}, \citenamefont {{Campos}}, \citenamefont {{Carnero Rosell}}, \citenamefont {{Carrasco Kind}}, \citenamefont {{Cawthon}}, \citenamefont {{Chen}}, \citenamefont {{Choi}}, \citenamefont {{Cordero}}, \citenamefont {{Davis}}, \citenamefont {{Diehl}}, \citenamefont {{Doux}}, \citenamefont {{Drlica-Wagner}}, \citenamefont
  {{Eckert}}, \citenamefont {{Elvin-Poole}}, \citenamefont {{Everett}}, \citenamefont {{Fert{\'e}}}, \citenamefont {{Gatti}}, \citenamefont {{Giannini}}, \citenamefont {{Gruen}}, \citenamefont {{Gruendl}}, \citenamefont {{Harrison}}, \citenamefont {{Herner}}, \citenamefont {{Huff}}, \citenamefont {{Jarvis}}, \citenamefont {{Kuropatkin}}, \citenamefont {{Leget}}, \citenamefont {{MacCrann}}, \citenamefont {{McCullough}}, \citenamefont {{Myles}}, \citenamefont {{Navarro-Alsina}}, \citenamefont {{Porredon}}, \citenamefont {{Raveri}}, \citenamefont {{Rollins}}, \citenamefont {{Roodman}}, \citenamefont {{Ross}}, \citenamefont {{Rykoff}}, \citenamefont {{Secco}}, \citenamefont {{Sevilla-Noarbe}}, \citenamefont {{Shin}}, \citenamefont {{Troxel}}, \citenamefont {{Tutusaus}}, \citenamefont {{Varga}}, \citenamefont {{Yanny}}, \citenamefont {{Yin}}, \citenamefont {{Zhang}}, \citenamefont {{Zuntz}}, \citenamefont {{Abbott}}, \citenamefont {{Aguena}}, \citenamefont {{Allam}}, \citenamefont {{Andrade-Oliveira}},
  \citenamefont {{Annis}}, \citenamefont {{Bacon}}, \citenamefont {{Bertin}}, \citenamefont {{Brooks}}, \citenamefont {{Burke}}, \citenamefont {{Carretero}}, \citenamefont {{Castander}}, \citenamefont {{Costanzi}}, \citenamefont {{da Costa}}, \citenamefont {{Pereira}}, \citenamefont {{Desai}}, \citenamefont {{Dietrich}}, \citenamefont {{Doel}}, \citenamefont {{Evrard}}, \citenamefont {{Flaugher}}, \citenamefont {{Frieman}}, \citenamefont {{Garc{\'\i}a-Bellido}}, \citenamefont {{Gaztanaga}}, \citenamefont {{Gschwend}}, \citenamefont {{Gutierrez}}, \citenamefont {{Hinton}}, \citenamefont {{Hollowood}}, \citenamefont {{Honscheid}}, \citenamefont {{Hoyle}}, \citenamefont {{James}}, \citenamefont {{Kuehn}}, \citenamefont {{Lima}}, \citenamefont {{Maia}}, \citenamefont {{Marshall}}, \citenamefont {{Melchior}}, \citenamefont {{Menanteau}}, \citenamefont {{Miquel}}, \citenamefont {{Muir}}, \citenamefont {{Ogando}}, \citenamefont {{Paz-Chinch{\'o}n}}, \citenamefont {{Pieres}}, \citenamefont {{Sanchez}}, \citenamefont
  {{Serrano}}, \citenamefont {{Smith}}, \citenamefont {{Suchyta}}, \citenamefont {{Tarle}}, \citenamefont {{Thomas}}, \citenamefont {{To}}, \citenamefont {{Wilkinson}},\ and\ \citenamefont {{DES Collaboration}}}]{Zacharegkas21}%
  \BibitemOpen
  \bibfield  {author} {\bibinfo {author} {\bibfnamefont {G.}~\bibnamefont {{Zacharegkas}}}, \bibnamefont {et~al.},\ }\href {https://doi.org/10.1093/mnras/stab3155} {\bibfield  {journal} {\bibinfo  {journal} {\mnras}\ }\textbf {\bibinfo {volume} {509}},\ \bibinfo {pages} {3119} (\bibinfo {year} {2022})}\BibitemShut {NoStop}%
\bibitem [{\citenamefont {{Elvin-Poole}}\ \emph {et~al.}(2023)\citenamefont {{Elvin-Poole}}, \citenamefont {{MacCrann}}, \citenamefont {{Everett}}, \citenamefont {{Prat}}, \citenamefont {{Rykoff}}, \citenamefont {{De Vicente}}, \citenamefont {{Yanny}}, \citenamefont {{Herner}}, \citenamefont {{Fert{\'e}}}, \citenamefont {{Di Valentino}}, \citenamefont {{Choi}}, \citenamefont {{Burke}}, \citenamefont {{Sevilla-Noarbe}}, \citenamefont {{Alarcon}}, \citenamefont {{Alves}}, \citenamefont {{Amon}}, \citenamefont {{Andrade-Oliveira}}, \citenamefont {{Baxter}}, \citenamefont {{Bechtol}}, \citenamefont {{Becker}}, \citenamefont {{Bernstein}}, \citenamefont {{Blazek}}, \citenamefont {{Camacho}}, \citenamefont {{Campos}}, \citenamefont {{Carnero Rosell}}, \citenamefont {{Carrasco Kind}}, \citenamefont {{Cawthon}}, \citenamefont {{Chang}}, \citenamefont {{Chen}}, \citenamefont {{Cordero}}, \citenamefont {{Crocce}}, \citenamefont {{Davis}}, \citenamefont {{DeRose}}, \citenamefont {{Diehl}}, \citenamefont {{Dodelson}},
  \citenamefont {{Doux}}, \citenamefont {{Drlica-Wagner}}, \citenamefont {{Eckert}}, \citenamefont {{Eifler}}, \citenamefont {{Elsner}}, \citenamefont {{Fang}}, \citenamefont {{Fosalba}}, \citenamefont {{Friedrich}}, \citenamefont {{Gatti}}, \citenamefont {{Giannini}}, \citenamefont {{Gruen}}, \citenamefont {{Gruendl}}, \citenamefont {{Harrison}}, \citenamefont {{Hartley}}, \citenamefont {{Huang}}, \citenamefont {{Huff}}, \citenamefont {{Huterer}}, \citenamefont {{Krause}}, \citenamefont {{Kuropatkin}}, \citenamefont {{Leget}}, \citenamefont {{Lemos}}, \citenamefont {{Liddle}}, \citenamefont {{McCullough}}, \citenamefont {{Muir}}, \citenamefont {{Myles}}, \citenamefont {{Navarro-Alsina}}, \citenamefont {{Pandey}}, \citenamefont {{Park}}, \citenamefont {{Porredon}}, \citenamefont {{Raveri}}, \citenamefont {{Rodriguez-Monroy}}, \citenamefont {{Rollins}}, \citenamefont {{Roodman}}, \citenamefont {{Rosenfeld}}, \citenamefont {{Ross}}, \citenamefont {{S{\'a}nchez}}, \citenamefont {{Sanchez}}, \citenamefont
  {{Secco}}, \citenamefont {{Sheldon}}, \citenamefont {{Shin}}, \citenamefont {{Troxel}}, \citenamefont {{Tutusaus}}, \citenamefont {{Varga}}, \citenamefont {{Weaverdyck}}, \citenamefont {{Wechsler}}, \citenamefont {{Yin}}, \citenamefont {{Zhang}}, \citenamefont {{Zuntz}}, \citenamefont {{Aguena}}, \citenamefont {{Avila}}, \citenamefont {{Bacon}}, \citenamefont {{Bertin}}, \citenamefont {{Bocquet}}, \citenamefont {{Brooks}}, \citenamefont {{Garc{\'\i}a-Bellido}}, \citenamefont {{Honscheid}}, \citenamefont {{Jarvis}}, \citenamefont {{Li}}, \citenamefont {{Mena-Fern{\'a}ndez}}, \citenamefont {{To}}, \citenamefont {{Wilkinson}},\ and\ \citenamefont {{DES Collaboration}}}]{DESY3Magnification2x2pt}%
  \BibitemOpen
  \bibfield  {author} {\bibinfo {author} {\bibfnamefont {J.}~\bibnamefont {{Elvin-Poole}}}, \bibnamefont {et~al.},\ }\href {https://doi.org/10.1093/mnras/stad1594} {\bibfield  {journal} {\bibinfo  {journal} {\mnras}\ }\textbf {\bibinfo {volume} {523}},\ \bibinfo {pages} {3649} (\bibinfo {year} {2023})}\BibitemShut {NoStop}%
\bibitem [{\citenamefont {{Dvornik}}\ \emph {et~al.}(2023)\citenamefont {{Dvornik}}, \citenamefont {{Heymans}}, \citenamefont {{Asgari}}, \citenamefont {{Mahony}}, \citenamefont {{Joachimi}}, \citenamefont {{Bilicki}}, \citenamefont {{Chisari}}, \citenamefont {{Hildebrandt}}, \citenamefont {{Hoekstra}}, \citenamefont {{Johnston}}, \citenamefont {{Kuijken}}, \citenamefont {{Mead}}, \citenamefont {{Miyatake}}, \citenamefont {{Nishimichi}}, \citenamefont {{Reischke}}, \citenamefont {{Unruh}},\ and\ \citenamefont {{Wright}}}]{KiDS2x2pt+SMF}%
  \BibitemOpen
  \bibfield  {author} {\bibinfo {author} {\bibfnamefont {A.}~\bibnamefont {{Dvornik}}}, \bibnamefont {et~al.},\ }\href {https://doi.org/10.1051/0004-6361/202245158} {\bibfield  {journal} {\bibinfo  {journal} {\aap}\ }\textbf {\bibinfo {volume} {675}},\ \bibinfo {eid} {A189} (\bibinfo {year} {2023})}\BibitemShut {NoStop}%
\bibitem [{\citenamefont {{Sugiyama}}\ \emph {et~al.}(2023)\citenamefont {{Sugiyama}}, \citenamefont {{Miyatake}}, \citenamefont {{More}}, \citenamefont {{Li}}, \citenamefont {{Shirasaki}}, \citenamefont {{Takada}}, \citenamefont {{Kobayashi}}, \citenamefont {{Takahashi}}, \citenamefont {{Nishimichi}}, \citenamefont {{Nishizawa}}, \citenamefont {{Rau}}, \citenamefont {{Zhang}}, \citenamefont {{Dalal}}, \citenamefont {{Mandelbaum}}, \citenamefont {{Strauss}}, \citenamefont {{Hamana}}, \citenamefont {{Oguri}}, \citenamefont {{Osato}}, \citenamefont {{Kannawadi}}, \citenamefont {{Hsieh}}, \citenamefont {{Luo}}, \citenamefont {{Armstrong}}, \citenamefont {{Bosch}}, \citenamefont {{Komiyama}}, \citenamefont {{Lupton}}, \citenamefont {{Lust}}, \citenamefont {{Miyazaki}}, \citenamefont {{Murayama}}, \citenamefont {{Okura}}, \citenamefont {{Price}}, \citenamefont {{Tait}}, \citenamefont {{Tanaka}},\ and\ \citenamefont {{Wang}}}]{HSCY32x2pt}%
  \BibitemOpen
  \bibfield  {author} {\bibinfo {author} {\bibfnamefont {S.}~\bibnamefont {{Sugiyama}}}, \bibnamefont {et~al.},\ }\href {https://doi.org/10.1103/PhysRevD.108.123521} {\bibfield  {journal} {\bibinfo  {journal} {\prd}\ }\textbf {\bibinfo {volume} {108}},\ \bibinfo {eid} {123521} (\bibinfo {year} {2023})}\BibitemShut {NoStop}%
\bibitem [{\citenamefont {{Sunayama}}\ \emph {et~al.}(2020)\citenamefont {{Sunayama}}, \citenamefont {{Park}}, \citenamefont {{Takada}}, \citenamefont {{Kobayashi}}, \citenamefont {{Nishimichi}}, \citenamefont {{Kurita}}, \citenamefont {{More}}, \citenamefont {{Oguri}},\ and\ \citenamefont {{Osato}}}]{Sunayama20}%
  \BibitemOpen
  \bibfield  {author} {\bibinfo {author} {\bibfnamefont {T.}~\bibnamefont {{Sunayama}}}, \bibinfo {author} {\bibfnamefont {Y.}~\bibnamefont {{Park}}}, \bibinfo {author} {\bibfnamefont {M.}~\bibnamefont {{Takada}}}, \bibinfo {author} {\bibfnamefont {Y.}~\bibnamefont {{Kobayashi}}}, \bibinfo {author} {\bibfnamefont {T.}~\bibnamefont {{Nishimichi}}}, \bibinfo {author} {\bibfnamefont {T.}~\bibnamefont {{Kurita}}}, \bibinfo {author} {\bibfnamefont {S.}~\bibnamefont {{More}}}, \bibinfo {author} {\bibfnamefont {M.}~\bibnamefont {{Oguri}}},\ \bibnamefont {and}\ \bibinfo {author} {\bibfnamefont {K.}~\bibnamefont {{Osato}}},\ }\href {https://doi.org/10.1093/mnras/staa1646} {\bibfield  {journal} {\bibinfo  {journal} {\mnras}\ }\textbf {\bibinfo {volume} {496}},\ \bibinfo {pages} {4468} (\bibinfo {year} {2020})}\BibitemShut {NoStop}%
\bibitem [{\citenamefont {{Wu}}\ \emph {et~al.}(2022)\citenamefont {{Wu}}, \citenamefont {{Costanzi}}, \citenamefont {{To}}, \citenamefont {{Salcedo}}, \citenamefont {{Weinberg}}, \citenamefont {{Annis}}, \citenamefont {{Bocquet}}, \citenamefont {{da Silva Pereira}}, \citenamefont {{DeRose}}, \citenamefont {{Esteves}}, \citenamefont {{Farahi}}, \citenamefont {{Grandis}}, \citenamefont {{Rozo}}, \citenamefont {{Rykoff}}, \citenamefont {{Varga}}, \citenamefont {{Wechsler}}, \citenamefont {{Zeng}}, \citenamefont {{Zhang}}, \citenamefont {{Zhang}},\ and\ \citenamefont {{DES Collaboration}}}]{Wuetal2022}%
  \BibitemOpen
  \bibfield  {author} {\bibinfo {author} {\bibfnamefont {H.-Y.}\ \bibnamefont {{Wu}}}, \bibnamefont {et~al.},\ }\href {https://doi.org/10.1093/mnras/stac2048} {\bibfield  {journal} {\bibinfo  {journal} {\mnras}\ }\textbf {\bibinfo {volume} {515}},\ \bibinfo {pages} {4471} (\bibinfo {year} {2022})}\BibitemShut {NoStop}%
\bibitem [{\citenamefont {{Zhou}}\ \emph {et~al.}(2024)\citenamefont {{Zhou}}, \citenamefont {{Wu}}, \citenamefont {{Salcedo}}, \citenamefont {{Grandis}}, \citenamefont {{Jeltema}}, \citenamefont {{Leauthaud}}, \citenamefont {{Costanzi}}, \citenamefont {{Sunayama}}, \citenamefont {{Weinberg}}, \citenamefont {{Zhang}}, \citenamefont {{Rozo}}, \citenamefont {{To}}, \citenamefont {{Bocquet}}, \citenamefont {{Varga}},\ and\ \citenamefont {{Kwiecien}}}]{Zhouetal2024}%
  \BibitemOpen
  \bibfield  {author} {\bibinfo {author} {\bibfnamefont {C.}~\bibnamefont {{Zhou}}}, \bibnamefont {et~al.},\ }\href {https://doi.org/10.1103/PhysRevD.110.103508} {\bibfield  {journal} {\bibinfo  {journal} {\prd}\ }\textbf {\bibinfo {volume} {110}},\ \bibinfo {eid} {103508} (\bibinfo {year} {2024})}\BibitemShut {NoStop}%
\bibitem [{\citenamefont {{Salcedo}}\ \emph {et~al.}(2024)\citenamefont {{Salcedo}}, \citenamefont {{Wu}}, \citenamefont {{Rozo}}, \citenamefont {{Weinberg}}, \citenamefont {{To}}, \citenamefont {{Sunayama}},\ and\ \citenamefont {{Lee}}}]{Salcedoetal2024}%
  \BibitemOpen
  \bibfield  {author} {\bibinfo {author} {\bibfnamefont {A.~N.}\ \bibnamefont {{Salcedo}}}, \bibinfo {author} {\bibfnamefont {H.-Y.}\ \bibnamefont {{Wu}}}, \bibinfo {author} {\bibfnamefont {E.}~\bibnamefont {{Rozo}}}, \bibinfo {author} {\bibfnamefont {D.~H.}\ \bibnamefont {{Weinberg}}}, \bibinfo {author} {\bibfnamefont {C.-H.}\ \bibnamefont {{To}}}, \bibinfo {author} {\bibfnamefont {T.}~\bibnamefont {{Sunayama}}},\ \bibnamefont {and}\ \bibinfo {author} {\bibfnamefont {A.}~\bibnamefont {{Lee}}},\ }\href {https://doi.org/10.1103/PhysRevLett.133.221002} {\bibfield  {journal} {\bibinfo  {journal} {\prl}\ }\textbf {\bibinfo {volume} {133}},\ \bibinfo {eid} {221002} (\bibinfo {year} {2024})}\BibitemShut {NoStop}%
\bibitem [{\citenamefont {{Cao}}\ \emph {et~al.}(2025)\citenamefont {{Cao}}, \citenamefont {{Wu}}, \citenamefont {{Costanzi}}, \citenamefont {{Farahi}}, \citenamefont {{Grandis}}, \citenamefont {{Weinberg}}, \citenamefont {{Evrard}}, \citenamefont {{Rozo}}, \citenamefont {{Salcedo}}, \citenamefont {{To}}, \citenamefont {{Yang}}, \citenamefont {{Zhou}},\ and\ \citenamefont {{DES Collaboration}}}]{Cao25association}%
  \BibitemOpen
  \bibfield  {author} {\bibinfo {author} {\bibfnamefont {S.}~\bibnamefont {{Cao}}}, \bibnamefont {et~al.},\ }\href {https://doi.org/10.1103/r7tt-bzs7} {\bibfield  {journal} {\bibinfo  {journal} {\prd}\ }\textbf {\bibinfo {volume} {112}},\ \bibinfo {eid} {043517} (\bibinfo {year} {2025})}\BibitemShut {NoStop}%
\bibitem [{\citenamefont {{Nyarko Nde}}\ \emph {et~al.}(2025)\citenamefont {{Nyarko Nde}}, \citenamefont {{Wu}}, \citenamefont {{Cao}}, \citenamefont {{Muthoni Kamau}}, \citenamefont {{Tamosiunas}}, \citenamefont {{To}},\ and\ \citenamefont {{Zhou}}}]{Ndeetal2025}%
  \BibitemOpen
  \bibfield  {author} {\bibinfo {author} {\bibfnamefont {T.}~\bibnamefont {{Nyarko Nde}}}, \bibinfo {author} {\bibfnamefont {H.-Y.}\ \bibnamefont {{Wu}}}, \bibinfo {author} {\bibfnamefont {S.}~\bibnamefont {{Cao}}}, \bibinfo {author} {\bibfnamefont {G.}~\bibnamefont {{Muthoni Kamau}}}, \bibinfo {author} {\bibfnamefont {A.}~\bibnamefont {{Tamosiunas}}}, \bibinfo {author} {\bibfnamefont {C.-H.}\ \bibnamefont {{To}}},\ \bibnamefont {and}\ \bibinfo {author} {\bibfnamefont {C.}~\bibnamefont {{Zhou}}},\ }\href {https://doi.org/10.48550/arXiv.2510.00753} {\bibfield  {journal} {\bibinfo  {journal} {arXiv e-prints}\ ,\ \bibinfo {eid} {arXiv:2510.00753}} (\bibinfo {year} {2025})}\BibitemShut {NoStop}%
\bibitem [{\citenamefont {{Zeng}}\ \emph {et~al.}(2023)\citenamefont {{Zeng}}, \citenamefont {{Salcedo}}, \citenamefont {{Wu}},\ and\ \citenamefont {{Hirata}}}]{Zeng23}%
  \BibitemOpen
  \bibfield  {author} {\bibinfo {author} {\bibfnamefont {C.}~\bibnamefont {{Zeng}}}, \bibinfo {author} {\bibfnamefont {A.~N.}\ \bibnamefont {{Salcedo}}}, \bibinfo {author} {\bibfnamefont {H.-Y.}\ \bibnamefont {{Wu}}},\ \bibnamefont {and}\ \bibinfo {author} {\bibfnamefont {C.~M.}\ \bibnamefont {{Hirata}}},\ }\href {https://doi.org/10.1093/mnras/stad1649} {\bibfield  {journal} {\bibinfo  {journal} {\mnras}\ }\textbf {\bibinfo {volume} {523}},\ \bibinfo {pages} {4270} (\bibinfo {year} {2023})}\BibitemShut {NoStop}%
\bibitem [{\citenamefont {{Park}}\ \emph {et~al.}(2023)\citenamefont {{Park}}, \citenamefont {{Sunayama}}, \citenamefont {{Takada}}, \citenamefont {{Kobayashi}}, \citenamefont {{Miyatake}}, \citenamefont {{More}}, \citenamefont {{Nishimichi}},\ and\ \citenamefont {{Sugiyama}}}]{Parketal2023}%
  \BibitemOpen
  \bibfield  {author} {\bibinfo {author} {\bibfnamefont {Y.}~\bibnamefont {{Park}}}, \bibinfo {author} {\bibfnamefont {T.}~\bibnamefont {{Sunayama}}}, \bibinfo {author} {\bibfnamefont {M.}~\bibnamefont {{Takada}}}, \bibinfo {author} {\bibfnamefont {Y.}~\bibnamefont {{Kobayashi}}}, \bibinfo {author} {\bibfnamefont {H.}~\bibnamefont {{Miyatake}}}, \bibinfo {author} {\bibfnamefont {S.}~\bibnamefont {{More}}}, \bibinfo {author} {\bibfnamefont {T.}~\bibnamefont {{Nishimichi}}},\ \bibnamefont {and}\ \bibinfo {author} {\bibfnamefont {S.}~\bibnamefont {{Sugiyama}}},\ }\href {https://doi.org/10.1093/mnras/stac3410} {\bibfield  {journal} {\bibinfo  {journal} {\mnras}\ }\textbf {\bibinfo {volume} {518}},\ \bibinfo {pages} {5171} (\bibinfo {year} {2023})}\BibitemShut {NoStop}%
\bibitem [{\citenamefont {{Sunayama}}\ \emph {et~al.}(2024)\citenamefont {{Sunayama}}, \citenamefont {{Miyatake}}, \citenamefont {{Sugiyama}}, \citenamefont {{More}}, \citenamefont {{Li}}, \citenamefont {{Dalal}}, \citenamefont {{Rau}}, \citenamefont {{Shi}}, \citenamefont {{Chiu}}, \citenamefont {{Shirasaki}}, \citenamefont {{Zhang}},\ and\ \citenamefont {{Nishizawa}}}]{Sunayamaetal2024}%
  \BibitemOpen
  \bibfield  {author} {\bibinfo {author} {\bibfnamefont {T.}~\bibnamefont {{Sunayama}}}, \bibnamefont {et~al.},\ }\href {https://doi.org/10.1103/PhysRevD.110.083511} {\bibfield  {journal} {\bibinfo  {journal} {\prd}\ }\textbf {\bibinfo {volume} {110}},\ \bibinfo {eid} {083511} (\bibinfo {year} {2024})}\BibitemShut {NoStop}%
\bibitem [{\citenamefont {{Wibking}}\ \emph {et~al.}(2020)\citenamefont {{Wibking}}, \citenamefont {{Weinberg}}, \citenamefont {{Salcedo}}, \citenamefont {{Wu}}, \citenamefont {{Singh}}, \citenamefont {{Rodr{\'\i}guez-Torres}}, \citenamefont {{Garrison}},\ and\ \citenamefont {{Eisenstein}}}]{Wibkingetal2020}%
  \BibitemOpen
  \bibfield  {author} {\bibinfo {author} {\bibfnamefont {B.~D.}\ \bibnamefont {{Wibking}}}, \bibinfo {author} {\bibfnamefont {D.~H.}\ \bibnamefont {{Weinberg}}}, \bibinfo {author} {\bibfnamefont {A.~N.}\ \bibnamefont {{Salcedo}}}, \bibinfo {author} {\bibfnamefont {H.-Y.}\ \bibnamefont {{Wu}}}, \bibinfo {author} {\bibfnamefont {S.}~\bibnamefont {{Singh}}}, \bibinfo {author} {\bibfnamefont {S.}~\bibnamefont {{Rodr{\'\i}guez-Torres}}}, \bibinfo {author} {\bibfnamefont {L.~H.}\ \bibnamefont {{Garrison}}},\ \bibnamefont {and}\ \bibinfo {author} {\bibfnamefont {D.~J.}\ \bibnamefont {{Eisenstein}}},\ }\href {https://doi.org/10.1093/mnras/stz3423} {\bibfield  {journal} {\bibinfo  {journal} {\mnras}\ }\textbf {\bibinfo {volume} {492}},\ \bibinfo {pages} {2872} (\bibinfo {year} {2020})}\BibitemShut {NoStop}%
\bibitem [{\citenamefont {{Salcedo}}\ \emph {et~al.}(2022)\citenamefont {{Salcedo}}, \citenamefont {{Weinberg}}, \citenamefont {{Wu}},\ and\ \citenamefont {{Wibking}}}]{Salcedoetal2022}%
  \BibitemOpen
  \bibfield  {author} {\bibinfo {author} {\bibfnamefont {A.~N.}\ \bibnamefont {{Salcedo}}}, \bibinfo {author} {\bibfnamefont {D.~H.}\ \bibnamefont {{Weinberg}}}, \bibinfo {author} {\bibfnamefont {H.-Y.}\ \bibnamefont {{Wu}}},\ \bibnamefont {and}\ \bibinfo {author} {\bibfnamefont {B.~D.}\ \bibnamefont {{Wibking}}},\ }\href {https://doi.org/10.1093/mnras/stab3793} {\bibfield  {journal} {\bibinfo  {journal} {\mnras}\ }\textbf {\bibinfo {volume} {510}},\ \bibinfo {pages} {5376} (\bibinfo {year} {2022})}\BibitemShut {NoStop}%
\bibitem [{\citenamefont {{Lee}}\ \emph {et~al.}(2025)\citenamefont {{Lee}}, \citenamefont {{Wu}}, \citenamefont {{Salcedo}}, \citenamefont {{Sunayama}}, \citenamefont {{Costanzi}}, \citenamefont {{Myles}}, \citenamefont {{Cao}}, \citenamefont {{Rozo}}, \citenamefont {{To}}, \citenamefont {{Weinberg}}, \citenamefont {{Yang}},\ and\ \citenamefont {{Zhou}}}]{Leeetal2024}%
  \BibitemOpen
  \bibfield  {author} {\bibinfo {author} {\bibfnamefont {A.}~\bibnamefont {{Lee}}}, \bibnamefont {et~al.},\ }\href {https://doi.org/10.1103/PhysRevD.111.063502} {\bibfield  {journal} {\bibinfo  {journal} {\prd}\ }\textbf {\bibinfo {volume} {111}},\ \bibinfo {eid} {063502} (\bibinfo {year} {2025})}\BibitemShut {NoStop}%
\bibitem [{\citenamefont {{Umetsu}}(2020)}]{Umetsu20}%
  \BibitemOpen
  \bibfield  {author} {\bibinfo {author} {\bibfnamefont {K.}~\bibnamefont {{Umetsu}}},\ }\href {https://doi.org/10.1007/s00159-020-00129-w} {\bibfield  {journal} {\bibinfo  {journal} {\aapr}\ }\textbf {\bibinfo {volume} {28}},\ \bibinfo {eid} {7} (\bibinfo {year} {2020})}\BibitemShut {NoStop}%
\bibitem [{\citenamefont {{Sinha}}\ and\ \citenamefont {{Garrison}}(2020)}]{CORRFUNC1}%
  \BibitemOpen
  \bibfield  {author} {\bibinfo {author} {\bibfnamefont {M.}~\bibnamefont {{Sinha}}}\ \bibnamefont {and}\ \bibinfo {author} {\bibfnamefont {L.~H.}\ \bibnamefont {{Garrison}}},\ }\href {https://doi.org/10.1093/mnras/stz3157} {\bibfield  {journal} {\bibinfo  {journal} {\mnras}\ }\textbf {\bibinfo {volume} {491}},\ \bibinfo {pages} {3022} (\bibinfo {year} {2020})}\BibitemShut {NoStop}%
\bibitem [{\citenamefont {{Wu}}\ \emph {et~al.}(2019)\citenamefont {{Wu}}, \citenamefont {{Weinberg}}, \citenamefont {{Salcedo}}, \citenamefont {{Wibking}},\ and\ \citenamefont {{Zu}}}]{Wuetal2019}%
  \BibitemOpen
  \bibfield  {author} {\bibinfo {author} {\bibfnamefont {H.-Y.}\ \bibnamefont {{Wu}}}, \bibinfo {author} {\bibfnamefont {D.~H.}\ \bibnamefont {{Weinberg}}}, \bibinfo {author} {\bibfnamefont {A.~N.}\ \bibnamefont {{Salcedo}}}, \bibinfo {author} {\bibfnamefont {B.~D.}\ \bibnamefont {{Wibking}}},\ \bibnamefont {and}\ \bibinfo {author} {\bibfnamefont {Y.}~\bibnamefont {{Zu}}},\ }\href {https://doi.org/10.1093/mnras/stz2617} {\bibfield  {journal} {\bibinfo  {journal} {\mnras}\ }\textbf {\bibinfo {volume} {490}},\ \bibinfo {pages} {2606} (\bibinfo {year} {2019})}\BibitemShut {NoStop}%
\bibitem [{\citenamefont {{Maksimova}}\ \emph {et~al.}(2021)\citenamefont {{Maksimova}}, \citenamefont {{Garrison}}, \citenamefont {{Eisenstein}}, \citenamefont {{Hadzhiyska}}, \citenamefont {{Bose}},\ and\ \citenamefont {{Satterthwaite}}}]{ABACUSSUMMIT}%
  \BibitemOpen
  \bibfield  {author} {\bibinfo {author} {\bibfnamefont {N.~A.}\ \bibnamefont {{Maksimova}}}, \bibinfo {author} {\bibfnamefont {L.~H.}\ \bibnamefont {{Garrison}}}, \bibinfo {author} {\bibfnamefont {D.~J.}\ \bibnamefont {{Eisenstein}}}, \bibinfo {author} {\bibfnamefont {B.}~\bibnamefont {{Hadzhiyska}}}, \bibinfo {author} {\bibfnamefont {S.}~\bibnamefont {{Bose}}},\ \bibnamefont {and}\ \bibinfo {author} {\bibfnamefont {T.~P.}\ \bibnamefont {{Satterthwaite}}},\ }\href {https://doi.org/10.1093/mnras/stab2484} {\bibfield  {journal} {\bibinfo  {journal} {\mnras}\ }\textbf {\bibinfo {volume} {508}},\ \bibinfo {pages} {4017} (\bibinfo {year} {2021})}\BibitemShut {NoStop}%
\bibitem [{\citenamefont {{Hadzhiyska}}\ \emph {et~al.}(2022)\citenamefont {{Hadzhiyska}}, \citenamefont {{Eisenstein}}, \citenamefont {{Bose}}, \citenamefont {{Garrison}},\ and\ \citenamefont {{Maksimova}}}]{HadzhiyskaCOMPASO}%
  \BibitemOpen
  \bibfield  {author} {\bibinfo {author} {\bibfnamefont {B.}~\bibnamefont {{Hadzhiyska}}}, \bibinfo {author} {\bibfnamefont {D.}~\bibnamefont {{Eisenstein}}}, \bibinfo {author} {\bibfnamefont {S.}~\bibnamefont {{Bose}}}, \bibinfo {author} {\bibfnamefont {L.~H.}\ \bibnamefont {{Garrison}}},\ \bibnamefont {and}\ \bibinfo {author} {\bibfnamefont {N.}~\bibnamefont {{Maksimova}}},\ }\href {https://doi.org/10.1093/mnras/stab2980} {\bibfield  {journal} {\bibinfo  {journal} {\mnras}\ }\textbf {\bibinfo {volume} {509}},\ \bibinfo {pages} {501} (\bibinfo {year} {2022})}\BibitemShut {NoStop}%
\bibitem [{\citenamefont {{Bose}}\ \emph {et~al.}(2022)\citenamefont {{Bose}}, \citenamefont {{Eisenstein}}, \citenamefont {{Hadzhiyska}}, \citenamefont {{Garrison}},\ and\ \citenamefont {{Yuan}}}]{BoseMergerTrees}%
  \BibitemOpen
  \bibfield  {author} {\bibinfo {author} {\bibfnamefont {S.}~\bibnamefont {{Bose}}}, \bibinfo {author} {\bibfnamefont {D.~J.}\ \bibnamefont {{Eisenstein}}}, \bibinfo {author} {\bibfnamefont {B.}~\bibnamefont {{Hadzhiyska}}}, \bibinfo {author} {\bibfnamefont {L.~H.}\ \bibnamefont {{Garrison}}},\ \bibnamefont {and}\ \bibinfo {author} {\bibfnamefont {S.}~\bibnamefont {{Yuan}}},\ }\href {https://doi.org/10.1093/mnras/stac555} {\bibfield  {journal} {\bibinfo  {journal} {\mnras}\ }\textbf {\bibinfo {volume} {512}},\ \bibinfo {pages} {837} (\bibinfo {year} {2022})}\BibitemShut {NoStop}%
\bibitem [{\citenamefont {{Bryan}}\ and\ \citenamefont {{Norman}}(1998)}]{BryanNorman1998}%
  \BibitemOpen
  \bibfield  {author} {\bibinfo {author} {\bibfnamefont {G.~L.}\ \bibnamefont {{Bryan}}}\ \bibnamefont {and}\ \bibinfo {author} {\bibfnamefont {M.~L.}\ \bibnamefont {{Norman}}},\ }\href {https://doi.org/10.1086/305262} {\bibfield  {journal} {\bibinfo  {journal} {\apj}\ }\textbf {\bibinfo {volume} {495}},\ \bibinfo {pages} {80} (\bibinfo {year} {1998})}\BibitemShut {NoStop}%
\bibitem [{\citenamefont {{Zheng}}\ \emph {et~al.}(2007)\citenamefont {{Zheng}}, \citenamefont {{Coil}},\ and\ \citenamefont {{Zehavi}}}]{Zheng07}%
  \BibitemOpen
  \bibfield  {author} {\bibinfo {author} {\bibfnamefont {Z.}~\bibnamefont {{Zheng}}}, \bibinfo {author} {\bibfnamefont {A.~L.}\ \bibnamefont {{Coil}}},\ \bibnamefont {and}\ \bibinfo {author} {\bibfnamefont {I.}~\bibnamefont {{Zehavi}}},\ }\href {https://doi.org/10.1086/521074} {\bibfield  {journal} {\bibinfo  {journal} {\apj}\ }\textbf {\bibinfo {volume} {667}},\ \bibinfo {pages} {760} (\bibinfo {year} {2007})}\BibitemShut {NoStop}%
\bibitem [{\citenamefont {{Navarro}}\ \emph {et~al.}(1997)\citenamefont {{Navarro}}, \citenamefont {{Frenk}},\ and\ \citenamefont {{White}}}]{NFW1997}%
  \BibitemOpen
  \bibfield  {author} {\bibinfo {author} {\bibfnamefont {J.~F.}\ \bibnamefont {{Navarro}}}, \bibinfo {author} {\bibfnamefont {C.~S.}\ \bibnamefont {{Frenk}}},\ \bibnamefont {and}\ \bibinfo {author} {\bibfnamefont {S.~D.~M.}\ \bibnamefont {{White}}},\ }\href {https://doi.org/10.1086/304888} {\bibfield  {journal} {\bibinfo  {journal} {\apj}\ }\textbf {\bibinfo {volume} {490}},\ \bibinfo {pages} {493} (\bibinfo {year} {1997})}\BibitemShut {NoStop}%
\bibitem [{\citenamefont {{Rykoff}}\ \emph {et~al.}(2014)\citenamefont {{Rykoff}}, \citenamefont {{Rozo}}, \citenamefont {{Busha}}, \citenamefont {{Cunha}}, \citenamefont {{Finoguenov}}, \citenamefont {{Evrard}}, \citenamefont {{Hao}}, \citenamefont {{Koester}}, \citenamefont {{Leauthaud}}, \citenamefont {{Nord}}, \citenamefont {{Pierre}}, \citenamefont {{Reddick}}, \citenamefont {{Sadibekova}}, \citenamefont {{Sheldon}},\ and\ \citenamefont {{Wechsler}}}]{Rykoffetal2014}%
  \BibitemOpen
  \bibfield  {author} {\bibinfo {author} {\bibfnamefont {E.~S.}\ \bibnamefont {{Rykoff}}}, \bibnamefont {et~al.},\ }\href {https://doi.org/10.1088/0004-637X/785/2/104} {\bibfield  {journal} {\bibinfo  {journal} {\apj}\ }\textbf {\bibinfo {volume} {785}},\ \bibinfo {eid} {104} (\bibinfo {year} {2014})}\BibitemShut {NoStop}%
\bibitem [{\citenamefont {{Salcedo}}\ \emph {et~al.}(2025{\natexlab{b}})\citenamefont {{Salcedo}}, \citenamefont {{Rozo}}, \citenamefont {{Wu}}, \citenamefont {{Weinberg}}, \citenamefont {{Chiploonkar}}, \citenamefont {{To}}, \citenamefont {{Cao}}, \citenamefont {{Rykoff}}, \citenamefont {{Marcelina Gountanis}},\ and\ \citenamefont {{Zhou}}}]{Salcedo2025a}%
  \BibitemOpen
  \bibfield  {author} {\bibinfo {author} {\bibfnamefont {A.~N.}\ \bibnamefont {{Salcedo}}}, \bibinfo {author} {\bibfnamefont {E.}~\bibnamefont {{Rozo}}}, \bibinfo {author} {\bibfnamefont {H.-Y.}\ \bibnamefont {{Wu}}}, \bibinfo {author} {\bibfnamefont {D.~H.}\ \bibnamefont {{Weinberg}}}, \bibinfo {author} {\bibfnamefont {P.}~\bibnamefont {{Chiploonkar}}}, \bibinfo {author} {\bibfnamefont {C.-H.}\ \bibnamefont {{To}}}, \bibinfo {author} {\bibfnamefont {S.}~\bibnamefont {{Cao}}}, \bibinfo {author} {\bibfnamefont {E.~S.}\ \bibnamefont {{Rykoff}}}, \bibinfo {author} {\bibfnamefont {N.}~\bibnamefont {{Marcelina Gountanis}}},\ \bibnamefont {and}\ \bibinfo {author} {\bibfnamefont {C.}~\bibnamefont {{Zhou}}},\ }\href {https://doi.org/10.48550/arXiv.2510.25706} {\bibfield  {journal} {\bibinfo  {journal} {arXiv e-prints}\ ,\ \bibinfo {eid} {arXiv:2510.25706}} (\bibinfo {year} {2025}{\natexlab{b}})}\BibitemShut {NoStop}%
\bibitem [{\citenamefont {{Nygaard}}\ \emph {et~al.}(2024)\citenamefont {{Nygaard}}, \citenamefont {{Holm}}, \citenamefont {{Hannestad}},\ and\ \citenamefont {{Tram}}}]{Nygaardetal2024}%
  \BibitemOpen
  \bibfield  {author} {\bibinfo {author} {\bibfnamefont {A.}~\bibnamefont {{Nygaard}}}, \bibinfo {author} {\bibfnamefont {E.~B.}\ \bibnamefont {{Holm}}}, \bibinfo {author} {\bibfnamefont {S.}~\bibnamefont {{Hannestad}}},\ \bibnamefont {and}\ \bibinfo {author} {\bibfnamefont {T.}~\bibnamefont {{Tram}}},\ }\href {https://doi.org/10.1088/1475-7516/2024/10/073} {\bibfield  {journal} {\bibinfo  {journal} {\jcap}\ }\textbf {\bibinfo {volume} {2024}},\ \bibinfo {eid} {073} (\bibinfo {year} {2024})}\BibitemShut {NoStop}%
\bibitem [{\citenamefont {{Schaye}}\ \emph {et~al.}(2023)\citenamefont {{Schaye}}, \citenamefont {{Kugel}}, \citenamefont {{Schaller}}, \citenamefont {{Helly}}, \citenamefont {{Braspenning}}, \citenamefont {{Elbers}}, \citenamefont {{McCarthy}}, \citenamefont {{van Daalen}}, \citenamefont {{Vandenbroucke}}, \citenamefont {{Frenk}}, \citenamefont {{Kwan}}, \citenamefont {{Salcido}}, \citenamefont {{Bah{\'e}}}, \citenamefont {{Borrow}}, \citenamefont {{Chaikin}}, \citenamefont {{Hahn}}, \citenamefont {{Hu{\v{s}}ko}}, \citenamefont {{Jenkins}}, \citenamefont {{Lacey}},\ and\ \citenamefont {{Nobels}}}]{Schayeetal2023}%
  \BibitemOpen
  \bibfield  {author} {\bibinfo {author} {\bibfnamefont {J.}~\bibnamefont {{Schaye}}}, \bibnamefont {et~al.},\ }\href {https://doi.org/10.1093/mnras/stad2419} {\bibfield  {journal} {\bibinfo  {journal} {\mnras}\ }\textbf {\bibinfo {volume} {526}},\ \bibinfo {pages} {4978} (\bibinfo {year} {2023})}\BibitemShut {NoStop}%
\bibitem [{\citenamefont {{Kugel}}\ \emph {et~al.}(2023)\citenamefont {{Kugel}}, \citenamefont {{Schaye}}, \citenamefont {{Schaller}}, \citenamefont {{Helly}}, \citenamefont {{Braspenning}}, \citenamefont {{Elbers}}, \citenamefont {{Frenk}}, \citenamefont {{McCarthy}}, \citenamefont {{Kwan}}, \citenamefont {{Salcido}}, \citenamefont {{van Daalen}}, \citenamefont {{Vandenbroucke}}, \citenamefont {{Bah{\'e}}}, \citenamefont {{Borrow}}, \citenamefont {{Chaikin}}, \citenamefont {{Hu{\v{s}}ko}}, \citenamefont {{Jenkins}}, \citenamefont {{Lacey}}, \citenamefont {{Nobels}},\ and\ \citenamefont {{Vernon}}}]{Kugel23Flamingo}%
  \BibitemOpen
  \bibfield  {author} {\bibinfo {author} {\bibfnamefont {R.}~\bibnamefont {{Kugel}}}, \bibnamefont {et~al.},\ }\href {https://doi.org/10.1093/mnras/stad2540} {\bibfield  {journal} {\bibinfo  {journal} {\mnras}\ }\textbf {\bibinfo {volume} {526}},\ \bibinfo {pages} {6103} (\bibinfo {year} {2023})}\BibitemShut {NoStop}%
\bibitem [{\citenamefont {{Kugel}}\ \emph {et~al.}(2025)\citenamefont {{Kugel}}, \citenamefont {{Schaye}}, \citenamefont {{Schaller}}, \citenamefont {{Forouhar Moreno}},\ and\ \citenamefont {{McGibbon}}}]{Kugel25FlamingoCluster}%
  \BibitemOpen
  \bibfield  {author} {\bibinfo {author} {\bibfnamefont {R.}~\bibnamefont {{Kugel}}}, \bibinfo {author} {\bibfnamefont {J.}~\bibnamefont {{Schaye}}}, \bibinfo {author} {\bibfnamefont {M.}~\bibnamefont {{Schaller}}}, \bibinfo {author} {\bibfnamefont {V.~J.}\ \bibnamefont {{Forouhar Moreno}}},\ \bibnamefont {and}\ \bibinfo {author} {\bibfnamefont {R.~J.}\ \bibnamefont {{McGibbon}}},\ }\href {https://doi.org/10.1093/mnras/staf111} {\bibfield  {journal} {\bibinfo  {journal} {\mnras}\ }\textbf {\bibinfo {volume} {537}},\ \bibinfo {pages} {2179} (\bibinfo {year} {2025})}\BibitemShut {NoStop}%
\bibitem [{\citenamefont {{Han}}\ \emph {et~al.}(2012)\citenamefont {{Han}}, \citenamefont {{Jing}}, \citenamefont {{Wang}},\ and\ \citenamefont {{Wang}}}]{Hanetal2012}%
  \BibitemOpen
  \bibfield  {author} {\bibinfo {author} {\bibfnamefont {J.}~\bibnamefont {{Han}}}, \bibinfo {author} {\bibfnamefont {Y.~P.}\ \bibnamefont {{Jing}}}, \bibinfo {author} {\bibfnamefont {H.}~\bibnamefont {{Wang}}},\ \bibnamefont {and}\ \bibinfo {author} {\bibfnamefont {W.}~\bibnamefont {{Wang}}},\ }\href {https://doi.org/10.1111/j.1365-2966.2012.22111.x} {\bibfield  {journal} {\bibinfo  {journal} {\mnras}\ }\textbf {\bibinfo {volume} {427}},\ \bibinfo {pages} {2437} (\bibinfo {year} {2012})}\BibitemShut {NoStop}%
\bibitem [{\citenamefont {{Forouhar Moreno}}\ \emph {et~al.}(2025)\citenamefont {{Forouhar Moreno}}, \citenamefont {{Helly}}, \citenamefont {{McGibbon}}, \citenamefont {{Schaye}}, \citenamefont {{Schaller}}, \citenamefont {{Han}}, \citenamefont {{Kugel}},\ and\ \citenamefont {{Bah{\'e}}}}]{Moreno2025HBT}%
  \BibitemOpen
  \bibfield  {author} {\bibinfo {author} {\bibfnamefont {V.~J.}\ \bibnamefont {{Forouhar Moreno}}}, \bibinfo {author} {\bibfnamefont {J.}~\bibnamefont {{Helly}}}, \bibinfo {author} {\bibfnamefont {R.}~\bibnamefont {{McGibbon}}}, \bibinfo {author} {\bibfnamefont {J.}~\bibnamefont {{Schaye}}}, \bibinfo {author} {\bibfnamefont {M.}~\bibnamefont {{Schaller}}}, \bibinfo {author} {\bibfnamefont {J.}~\bibnamefont {{Han}}}, \bibinfo {author} {\bibfnamefont {R.}~\bibnamefont {{Kugel}}},\ \bibnamefont {and}\ \bibinfo {author} {\bibfnamefont {Y.~M.}\ \bibnamefont {{Bah{\'e}}}},\ }\href {https://doi.org/10.1093/mnras/staf1478} {\bibfield  {journal} {\bibinfo  {journal} {\mnras}\ }\textbf {\bibinfo {volume} {543}},\ \bibinfo {pages} {1339} (\bibinfo {year} {2025})}\BibitemShut {NoStop}%
\bibitem [{\citenamefont {{Rozo}}\ \emph {et~al.}(2016)\citenamefont {{Rozo}}, \citenamefont {{Rykoff}}, \citenamefont {{Abate}}, \citenamefont {{Bonnett}}, \citenamefont {{Crocce}}, \citenamefont {{Davis}}, \citenamefont {{Hoyle}}, \citenamefont {{Leistedt}}, \citenamefont {{Peiris}},\ and\ \citenamefont {{Wechsler et al}}}]{Rozoetal2016}%
  \BibitemOpen
  \bibfield  {author} {\bibinfo {author} {\bibfnamefont {E.}~\bibnamefont {{Rozo}}}, \bibinfo {author} {\bibfnamefont {E.~S.}\ \bibnamefont {{Rykoff}}}, \bibinfo {author} {\bibfnamefont {A.}~\bibnamefont {{Abate}}}, \bibinfo {author} {\bibfnamefont {C.}~\bibnamefont {{Bonnett}}}, \bibinfo {author} {\bibfnamefont {M.}~\bibnamefont {{Crocce}}}, \bibinfo {author} {\bibfnamefont {C.}~\bibnamefont {{Davis}}}, \bibinfo {author} {\bibfnamefont {B.}~\bibnamefont {{Hoyle}}}, \bibinfo {author} {\bibfnamefont {B.}~\bibnamefont {{Leistedt}}}, \bibinfo {author} {\bibfnamefont {H.~V.}\ \bibnamefont {{Peiris}}},\ \bibnamefont {and}\ \bibinfo {author} {\bibfnamefont {R.~H.}\ \bibnamefont {{Wechsler et al}}},\ }\href {https://doi.org/10.1093/mnras/stw1281} {\bibfield  {journal} {\bibinfo  {journal} {\mnras}\ }\textbf {\bibinfo {volume} {461}},\ \bibinfo {pages} {1431} (\bibinfo {year} {2016})}\BibitemShut {NoStop}%
\bibitem [{\citenamefont {{Foreman-Mackey}}\ \emph {et~al.}(2013)\citenamefont {{Foreman-Mackey}}, \citenamefont {{Hogg}}, \citenamefont {{Lang}},\ and\ \citenamefont {{Goodman}}}]{emcee}%
  \BibitemOpen
  \bibfield  {author} {\bibinfo {author} {\bibfnamefont {D.}~\bibnamefont {{Foreman-Mackey}}}, \bibinfo {author} {\bibfnamefont {D.~W.}\ \bibnamefont {{Hogg}}}, \bibinfo {author} {\bibfnamefont {D.}~\bibnamefont {{Lang}}},\ \bibnamefont {and}\ \bibinfo {author} {\bibfnamefont {J.}~\bibnamefont {{Goodman}}},\ }\href {https://doi.org/10.1086/670067} {\bibfield  {journal} {\bibinfo  {journal} {Publ. Astron. Soc. Pac.}\ }\textbf {\bibinfo {volume} {125}},\ \bibinfo {pages} {306} (\bibinfo {year} {2013})}\BibitemShut {NoStop}%
\bibitem [{\citenamefont {{Leauthaud}}\ \emph {et~al.}(2017)\citenamefont {{Leauthaud}}, \citenamefont {{Saito}}, \citenamefont {{Hilbert}}, \citenamefont {{Barreira}}, \citenamefont {{More}}, \citenamefont {{White}}, \citenamefont {{Alam}}, \citenamefont {{Behroozi}}, \citenamefont {{Bundy}}, \citenamefont {{Coupon}}, \citenamefont {{Erben}}, \citenamefont {{Heymans}}, \citenamefont {{Hildebrandt}}, \citenamefont {{Mandelbaum}}, \citenamefont {{Miller}}, \citenamefont {{Moraes}}, \citenamefont {{Pereira}}, \citenamefont {{Rodr{\'\i}guez-Torres}}, \citenamefont {{Schmidt}}, \citenamefont {{Shan}}, \citenamefont {{Viel}},\ and\ \citenamefont {{Villaescusa-Navarro}}}]{Leauthaud17}%
  \BibitemOpen
  \bibfield  {author} {\bibinfo {author} {\bibfnamefont {A.}~\bibnamefont {{Leauthaud}}}, \bibnamefont {et~al.},\ }\href {https://doi.org/10.1093/mnras/stx258} {\bibfield  {journal} {\bibinfo  {journal} {\mnras}\ }\textbf {\bibinfo {volume} {467}},\ \bibinfo {pages} {3024} (\bibinfo {year} {2017})}\BibitemShut {NoStop}%
\bibitem [{\citenamefont {{Lange}}\ \emph {et~al.}(2019)\citenamefont {{Lange}}, \citenamefont {{Yang}}, \citenamefont {{Guo}}, \citenamefont {{Luo}},\ and\ \citenamefont {{van den Bosch}}}]{Lange19}%
  \BibitemOpen
  \bibfield  {author} {\bibinfo {author} {\bibfnamefont {J.~U.}\ \bibnamefont {{Lange}}}, \bibinfo {author} {\bibfnamefont {X.}~\bibnamefont {{Yang}}}, \bibinfo {author} {\bibfnamefont {H.}~\bibnamefont {{Guo}}}, \bibinfo {author} {\bibfnamefont {W.}~\bibnamefont {{Luo}}},\ \bibnamefont {and}\ \bibinfo {author} {\bibfnamefont {F.~C.}\ \bibnamefont {{van den Bosch}}},\ }\href {https://doi.org/10.1093/mnras/stz2124} {\bibfield  {journal} {\bibinfo  {journal} {\mnras}\ }\textbf {\bibinfo {volume} {488}},\ \bibinfo {pages} {5771} (\bibinfo {year} {2019})}\BibitemShut {NoStop}%
\bibitem [{\citenamefont {{Krause}}\ \emph {et~al.}(2021)\citenamefont {{Krause}}, \citenamefont {{Fang}}, \citenamefont {{Pandey}}, \citenamefont {{Secco}}, \citenamefont {{Alves}}, \citenamefont {{Huang}}, \citenamefont {{Blazek}}, \citenamefont {{Prat}}, \citenamefont {{Zuntz}}, \citenamefont {{Eifler}}, \citenamefont {{MacCrann}}, \citenamefont {{DeRose}}, \citenamefont {{Crocce}}, \citenamefont {{Porredon}}, \citenamefont {{Jain}}, \citenamefont {{Troxel}}, \citenamefont {{Dodelson}}, \citenamefont {{Huterer}}, \citenamefont {{Liddle}}, \citenamefont {{Leonard}} \emph {et~al.}}]{Krause21}%
  \BibitemOpen
  \bibfield  {author} {\bibinfo {author} {\bibfnamefont {E.}~\bibnamefont {{Krause}}}, \bibnamefont {et~al.},\ }\href {https://doi.org/10.48550/arXiv.2105.13548} {\bibfield  {journal} {\bibinfo  {journal} {arXiv e-prints}\ ,\ \bibinfo {eid} {arXiv:2105.13548}} (\bibinfo {year} {2021})}\BibitemShut {NoStop}%
\bibitem [{\citenamefont {{Myles}}\ \emph {et~al.}(2021)\citenamefont {{Myles}}, \citenamefont {{Alarcon}}, \citenamefont {{Amon}}, \citenamefont {{S{\'a}nchez}}, \citenamefont {{Everett}}, \citenamefont {{DeRose}}, \citenamefont {{McCullough}}, \citenamefont {{Gruen}}, \citenamefont {{Bernstein}}, \citenamefont {{Troxel}}, \citenamefont {{Dodelson}}, \citenamefont {{Campos}}, \citenamefont {{MacCrann}}, \citenamefont {{Yin}}, \citenamefont {{Raveri}}, \citenamefont {{Amara}}, \citenamefont {{Becker}}, \citenamefont {{Choi}}, \citenamefont {{Cordero}}, \citenamefont {{Eckert}}, \citenamefont {{Gatti}}, \citenamefont {{Giannini}}, \citenamefont {{Gschwend}}, \citenamefont {{Gruendl}}, \citenamefont {{Harrison}}, \citenamefont {{Hartley}}, \citenamefont {{Huff}}, \citenamefont {{Kuropatkin}}, \citenamefont {{Lin}}, \citenamefont {{Masters}}, \citenamefont {{Miquel}}, \citenamefont {{Prat}}, \citenamefont {{Roodman}}, \citenamefont {{Rykoff}}, \citenamefont {{Sevilla-Noarbe}}, \citenamefont {{Sheldon}},
  \citenamefont {{Wechsler}}, \citenamefont {{Yanny}}, \citenamefont {{Abbott}}, \citenamefont {{Aguena}}, \citenamefont {{Allam}}, \citenamefont {{Annis}}, \citenamefont {{Bacon}}, \citenamefont {{Bertin}}, \citenamefont {{Bhargava}}, \citenamefont {{Bridle}}, \citenamefont {{Brooks}}, \citenamefont {{Burke}}, \citenamefont {{Carnero Rosell}}, \citenamefont {{Carrasco Kind}}, \citenamefont {{Carretero}}, \citenamefont {{Castander}}, \citenamefont {{Conselice}}, \citenamefont {{Costanzi}}, \citenamefont {{Crocce}}, \citenamefont {{da Costa}}, \citenamefont {{Pereira}}, \citenamefont {{Desai}}, \citenamefont {{Diehl}}, \citenamefont {{Eifler}}, \citenamefont {{Elvin-Poole}}, \citenamefont {{Evrard}}, \citenamefont {{Ferrero}}, \citenamefont {{Fert{\'e}}}, \citenamefont {{Flaugher}}, \citenamefont {{Fosalba}}, \citenamefont {{Frieman}}, \citenamefont {{Garc{\'\i}a-Bellido}}, \citenamefont {{Gaztanaga}}, \citenamefont {{Giannantonio}}, \citenamefont {{Hinton}}, \citenamefont {{Hollowood}}, \citenamefont
  {{Honscheid}}, \citenamefont {{Hoyle}}, \citenamefont {{Huterer}}, \citenamefont {{James}}, \citenamefont {{Krause}}, \citenamefont {{Kuehn}}, \citenamefont {{Lahav}}, \citenamefont {{Lima}}, \citenamefont {{Maia}}, \citenamefont {{Marshall}}, \citenamefont {{Martini}}, \citenamefont {{Melchior}}, \citenamefont {{Menanteau}}, \citenamefont {{Mohr}}, \citenamefont {{Morgan}}, \citenamefont {{Muir}}, \citenamefont {{Ogando}}, \citenamefont {{Palmese}}, \citenamefont {{Paz-Chinch{\'o}n}}, \citenamefont {{Plazas}}, \citenamefont {{Rodriguez-Monroy}}, \citenamefont {{Samuroff}}, \citenamefont {{Sanchez}}, \citenamefont {{Scarpine}}, \citenamefont {{Secco}}, \citenamefont {{Serrano}}, \citenamefont {{Smith}}, \citenamefont {{Soares-Santos}}, \citenamefont {{Suchyta}}, \citenamefont {{Swanson}}, \citenamefont {{Tarle}}, \citenamefont {{Thomas}}, \citenamefont {{To}}, \citenamefont {{Varga}}, \citenamefont {{Weller}},\ and\ \citenamefont {{Wester}}}]{Myles21SOMPZ}%
  \BibitemOpen
  \bibfield  {author} {\bibinfo {author} {\bibfnamefont {J.}~\bibnamefont {{Myles}}}, \bibnamefont {et~al.},\ }\href {https://doi.org/10.1093/mnras/stab1515} {\bibfield  {journal} {\bibinfo  {journal} {\mnras}\ }\textbf {\bibinfo {volume} {505}},\ \bibinfo {pages} {4249} (\bibinfo {year} {2021})}\BibitemShut {NoStop}%
\bibitem [{\citenamefont {{Gatti}}\ \emph {et~al.}(2021)\citenamefont {{Gatti}}, \citenamefont {{Sheldon}}, \citenamefont {{Amon}}, \citenamefont {{Becker}}, \citenamefont {{Troxel}}, \citenamefont {{Choi}}, \citenamefont {{Doux}}, \citenamefont {{MacCrann}}, \citenamefont {{Navarro-Alsina}}, \citenamefont {{Harrison}}, \citenamefont {{Gruen}}, \citenamefont {{Bernstein}}, \citenamefont {{Jarvis}}, \citenamefont {{Secco}}, \citenamefont {{Fert{\'e}}}, \citenamefont {{Shin}}, \citenamefont {{McCullough}}, \citenamefont {{Rollins}}, \citenamefont {{Chen}}, \citenamefont {{Chang}}, \citenamefont {{Pandey}}, \citenamefont {{Tutusaus}}, \citenamefont {{Prat}}, \citenamefont {{Elvin-Poole}}, \citenamefont {{Sanchez}}, \citenamefont {{Plazas}}, \citenamefont {{Roodman}}, \citenamefont {{Zuntz}}, \citenamefont {{Abbott}}, \citenamefont {{Aguena}}, \citenamefont {{Allam}}, \citenamefont {{Annis}}, \citenamefont {{Avila}}, \citenamefont {{Bacon}}, \citenamefont {{Bertin}}, \citenamefont {{Bhargava}}, \citenamefont
  {{Brooks}}, \citenamefont {{Burke}} \emph {et~al.}}]{Gatti21shear}%
  \BibitemOpen
  \bibfield  {author} {\bibinfo {author} {\bibfnamefont {M.}~\bibnamefont {{Gatti}}}, \bibnamefont {et~al.},\ }\href {https://doi.org/10.1093/mnras/stab918} {\bibfield  {journal} {\bibinfo  {journal} {\mnras}\ }\textbf {\bibinfo {volume} {504}},\ \bibinfo {pages} {4312} (\bibinfo {year} {2021})}\BibitemShut {NoStop}%
\bibitem [{\citenamefont {{Yamamoto}}\ \emph {et~al.}(2025)\citenamefont {{Yamamoto}}, \citenamefont {{Becker}}, \citenamefont {{Sheldon}}, \citenamefont {{Jarvis}}, \citenamefont {{Gruendl}}, \citenamefont {{Menanteau}}, \citenamefont {{Rykoff}}, \citenamefont {{Mau}}, \citenamefont {{Schutt}}, \citenamefont {{Gatti}}, \citenamefont {{Troxel}}, \citenamefont {{Amon}}, \citenamefont {{Anbajagane}}, \citenamefont {{Bernstein}}, \citenamefont {{Gruen}}, \citenamefont {{Huff}}, \citenamefont {{Tabbutt}}, \citenamefont {{Tong}}, \citenamefont {{Yanny}}, \citenamefont {{Abbott}}, \citenamefont {{Aguena}}, \citenamefont {{Alarcon}}, \citenamefont {{Andrade-Oliveira}}, \citenamefont {{Bechtol}}, \citenamefont {{Blazek}}, \citenamefont {{Brooks}}, \citenamefont {{Rosell}}, \citenamefont {{Carretero}}, \citenamefont {{Chang}}, \citenamefont {{Choi}}, \citenamefont {{Costanzi}}, \citenamefont {{Crocce}}, \citenamefont {{da Costa}}, \citenamefont {{Davis}}, \citenamefont {{De Vicente}}, \citenamefont {{Desai}},
  \citenamefont {{Diehl}}, \citenamefont {{Dodelson}}, \citenamefont {{Doel}}, \citenamefont {{Doux}}, \citenamefont {{Drlica-Wagner}}, \citenamefont {{Fert{\'e}}}, \citenamefont {{Flaugher}}, \citenamefont {{Fosalba}}, \citenamefont {{Frieman}}, \citenamefont {{Garc{\'\i}a-Bellido}}, \citenamefont {{Gaztanaga}}, \citenamefont {{Giannini}}, \citenamefont {{Gutierrez}}, \citenamefont {{Hartley}}, \citenamefont {{Herner}}, \citenamefont {{Hinton}}, \citenamefont {{Hollowood}}, \citenamefont {{Honscheid}}, \citenamefont {{Huterer}}, \citenamefont {{Krause}}, \citenamefont {{Kuehn}}, \citenamefont {{Lahav}}, \citenamefont {{Lima}}, \citenamefont {{Marshall}}, \citenamefont {{Mena-Fern{\'a}ndez}}, \citenamefont {{Miquel}}, \citenamefont {{Mohr}}, \citenamefont {{Muir}}, \citenamefont {{Myles}}, \citenamefont {{Ogando}}, \citenamefont {{Pieres}}, \citenamefont {{Malag{\'o}n}}, \citenamefont {{Porredon}}, \citenamefont {{Prat}}, \citenamefont {{Raveri}}, \citenamefont {{Rodriguez-Monroy}}, \citenamefont {{Roodman}},
  \citenamefont {{Samuroff}}, \citenamefont {{Sanchez}}, \citenamefont {{Sanchez Cid}}, \citenamefont {{Scarpine}}, \citenamefont {{Sevilla-Noarbe}}, \citenamefont {{Smith}}, \citenamefont {{Soares-Santos}}, \citenamefont {{Suchyta}}, \citenamefont {{Tarle}}, \citenamefont {{Vikram}}, \citenamefont {{Weaverdyck}}, \citenamefont {{Wiseman}},\ and\ \citenamefont {{Zhang}}}]{Yamamoto25}%
  \BibitemOpen
  \bibfield  {author} {\bibinfo {author} {\bibfnamefont {M.}~\bibnamefont {{Yamamoto}}}, \bibnamefont {et~al.},\ }\href {https://doi.org/10.1093/mnras/staf1661} {\bibfield  {journal} {\bibinfo  {journal} {\mnras}\ }\textbf {\bibinfo {volume} {543}},\ \bibinfo {pages} {4156} (\bibinfo {year} {2025})}\BibitemShut {NoStop}%
\end{thebibliography}%


\end{document}